\documentclass[11pt]{article}

\usepackage[preprint]{acl}

\usepackage{times}
\usepackage{latexsym}

\usepackage[T1]{fontenc}

\usepackage[utf8]{inputenc}

\usepackage{microtype}

\usepackage{inconsolata}

\usepackage{graphicx}
\graphicspath{{figures/}}

\usepackage{amsmath}    
\usepackage{amssymb}    

\usepackage{booktabs}
\usepackage{multirow}
\usepackage{array}
\usepackage{makecell}
\usepackage{enumitem} 
\usepackage{float}

\usepackage[ruled,vlined,linesnumbered]{algorithm2e}
\usepackage[most]{tcolorbox}

\usepackage{tabularx}

\usepackage{adjustbox}

\newcolumntype{Y}{>{\raggedright\arraybackslash}X}
\newcolumntype{Z}{>{\centering\arraybackslash}X}
\newcolumntype{L}[1]{>{\raggedright\arraybackslash}p{#1}}
\newcolumntype{C}[1]{>{\centering\arraybackslash}p{#1}}
\usepackage{xspace} 
\newcommand{\fitmath}[1]{\adjustbox{max width=\linewidth}{\(\displaystyle #1\)}}
\newcommand{\ourmethod}{{\fontfamily{lmtt}\selectfont \textbf{CtrlBench-Rec}}\xspace}

\title{Can We Steer the Black-Box? Towards Controllability-Centric Evaluation of Recommender Systems with Collaborative Agents}

\author{
  \textbf{Jiwen Zhou}\textsuperscript{1,2}
  \thanks{Equal contribution.}
  \and
  \textbf{Xiang Liu}\textsuperscript{1,2}\footnotemark[1]
  \and
  \textbf{Mingming Li}\textsuperscript{1}
  \thanks{Corresponding authors.}
  \and
  \textbf{Pengbo Mo}\textsuperscript{1,2}
  \and \\
  \textbf{Jiao Dai}\textsuperscript{1}
  \and
  \textbf{Honglei Lv}\textsuperscript{1}\footnotemark[2]
  \and
  \textbf{Jizhong Han}\textsuperscript{1}
  \and
  \textbf{Songlin Hu}\textsuperscript{1,2}
  \\
  \textsuperscript{1}Institute of Information Engineering,
  Chinese Academy of Sciences, China
  \\
  \textsuperscript{2}School of Cyber Security,
  University of Chinese Academy of Sciences, China
  \\
  \small{
    \texttt{\{zhoujiwen, liuxiang, limingming, mopengbo, daijiao,
    lvhonglei, hanjizhong, husonglin\}@iie.ac.cn}
  }
}

\begin{document}

\maketitle

\begin{abstract}
Recommender systems operate as Black-Boxes, leaving users and regulators unable to steer their outputs toward specific intentions or audit their behavior. This lack of controllability, defined as the system's ability to respond to explicit guidance, remains an unaddressed dimension in existing evaluation paradigms. To fill this gap, we propose CtrlBench-Rec, a collaborative multi-agent framework for systematic assessment of controllability. We formalize three fundamental tasks: target content discovery, interest profile shaping, and popularity bias mitigation, which together measure steerability from explicit commands to implicit representation steering and finally to overcoming algorithmic biases.
Extensive experiments on real-world datasets and multiple recommendation models demonstrate that our framework effectively quantifies controllability and exposes critical system bottlenecks, most notably persistent resistance to guiding long tail content.  \ourmethod provides the first standardized toolkit for controllable recommendation research, algorithmic auditing, and user empowerment.  Our code is released on  {\url{https://github.com/caskcsg/CtrlBenchRec}}.

\end{abstract}

\section{Introduction}
Recommender systems shape how users access and consume information, yet their complex Black-Box nature obscures the logic driving recommendations. This lack of transparency leaves users unable to comprehend or guide recommendations  \cite{zhang2020explainable}, raising urgent concerns. Regulators demand audit tools to detect bias or harmful content  \cite{chen2023bias}, while users seek to break information bubbles or guide exploration toward specific goals, such as deep engagement with niche topics or temporary avoidance of certain content  \cite{wang2022user,lukoff2021design,lukoff2023switchtube}. These concerns highlight a fundamental yet neglected property of recommender systems: controllability, the capacity to respond to explicit user intents and external guidance.

Achieving targeted guidance for Black-Box recommender systems faces a fundamental effic    iency bottleneck. Unlike search systems with explicit queries, recommenders rely on implicit behavior sequences like clicks and ratings  \cite{si2023search,fan2023adversarial}, so guidance must inject crafted behavior sequences to reshape the user profile  \cite{hou2024large,liu2025novel}. Each injection is costly, making guidance a resource-constrained sequential decision problem where inefficient exploration exhausts budgets. Multi-agent collaboration can accelerate discovery  \cite{wang2024multi}, but uncoordinated scaling merely shifts the cost burden  \cite{cai2025agentbalance,yang2026understanding}. The core challenge is distilling many agents' broad exploration into a few agents' precise control.


The fundamental deficit in current recommendation research stems from the absence of a standardized framework for controllability, a methodological void that has led to persistent fragmentation and hindered cross-study comparability \cite{zhang2026transparent}. Existing approaches fall short in addressing this gap \cite{liu2025recoworld}. On one hand, traditional benchmarks heavily prioritize predictive accuracy and rely on static post-hoc analyses (e.g., interpretability and fairness), lacking dynamic interaction with live systems  \cite{zhao2021recbole,jadon2024comprehensive}. On the other hand, methods that train a single agent via reinforcement learning for fixed long-term objectives struggle to accommodate diverse, immediate control goals under limited budgets  \cite{wang2025user,zhang2024agentcf}. Consequently, neither paradigm offers a rigorous, reproducible quantitative assessment of a system's responsiveness to external guidance, nor do they support efficient multi-agent co-evolution under resource constraints.

\begin{figure}[tp]
  \centering
  \includegraphics[width=\linewidth,keepaspectratio]{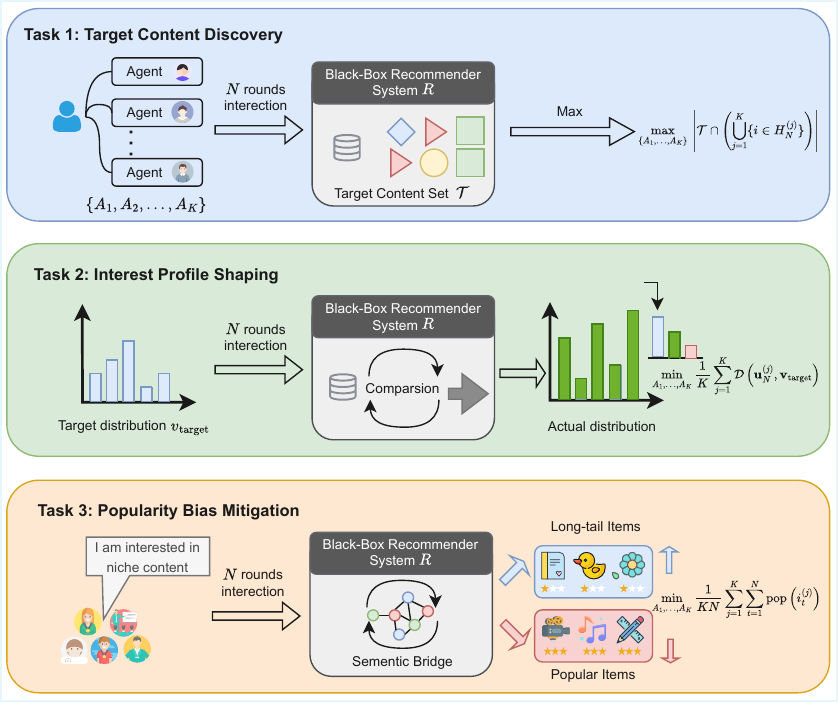}
  \caption{A taxonomy of controllability tasks for Black-Box recommender systems.}
  \label{tasks}
\end{figure}

To bridge this gap, we establish controllability as a core evaluative dimension and propose CtrlBench-Rec, the first multi-agent framework for systematic controllability assessment. We formalize three tasks in Figure~\ref{tasks} that progressively measure steerability: \textbf{Target Content Discovery} for responsiveness to explicit commands, \textbf{Interest Profile Shaping} for the ability to form a complex intended profile, and \textbf{Popularity Bias Mitigation} for the capacity to overcome inherent biases and uncover long tail content. To address the high cost of Black-Box interactions, \ourmethod introduces an evolutionary multi-agent fusion algorithm that evolves a diverse initial population into a compact set of specialized super probes via behavior analysis and functional clustering, significantly enhancing exploration efficiency. Our main contributions are threefold:

\begin{itemize}[leftmargin=*, nosep]
\item A multi-level controllability task suite that operationalizes external steerability from explicit goals to implicit representation shaping.
\item An evolutionary multi-agent framework that transforms evaluation into an optimizable probe cultivation process, achieving efficient Black-Box probing.
\item Extensive experiments on real-world datasets and multiple Black-Box models that validate the framework's effectiveness and reveal systemic controllability bottlenecks, most persistent resistance to guiding long-tail content.
\end{itemize}

\section{Problem Definition and Task Formalization}
We formalize controllability evaluation for Black-Box recommender systems. Key notations are summarized in Table~\ref{tab:notation}; full details, including formal optimization objectives, are deferred to Appendix~\ref{app:formulation}.

\subsection{ Preliminary Definitions}
Let  $\mathcal{I}$ be the universal item set, and $\mathcal{T}\subseteq \mathcal{I}$ a target content set. The Black-Box recommender system $\mathcal{R}$  takes a user's historical interaction sequence $H_{t}$ as input and returns a ranked recommendation list $L_t$. 
An agent probe  $A$ follows policy $\pi$ to  select actions $a_t$, (e.g., clicking an item), from
$L_t$. 
Each interaction incurs a cost, and the total number of rounds $N$  is strictly budgeted.  
A population of $K$ agents interacts independently, and evaluation aggregates their behavior histories.

\begin{figure*}[tp]
    \centering
    \includegraphics[scale=0.8]{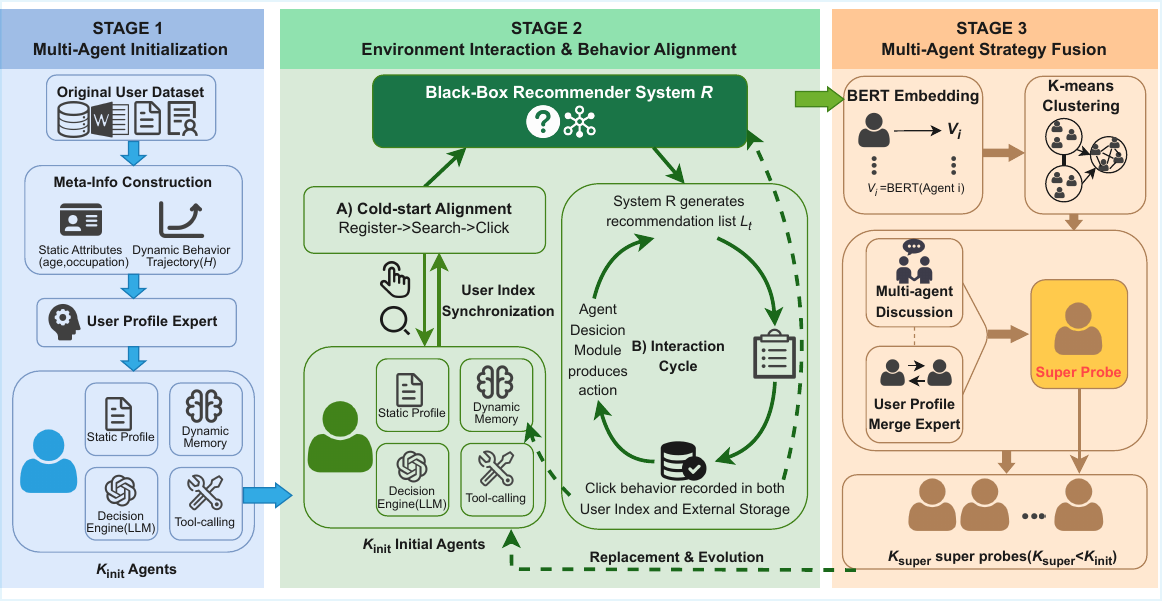}
    \caption{The overview of the evolutionary training process for the \ourmethod framework.}
    \label{fig:overview}
\end{figure*}

\subsection{Three Core Controllability Tasks}
We define three tasks that progressively measure steerability from explicit goals to implicit representation shaping and bias correction. Formal objective is shown in Appendix~\ref{app:t1}.

\textbf{Task 1: Target Content Discovery.}
This task evaluates whether the system can be steered to surface items from a predefined target set $\mathcal{T}$ within a limited budget. The goal is to maximize the number of unique discovered target items across all agents. 

\textbf{Task 2: Interest Profile Shaping.}
This task examines whether the system can be guided toward a desired multi-dimensional interest distribution $\mathbf{v}_{\text{target}}$. Since internal representations are unobservable, we measure input-output consistency by observing whether the recommended item distribution aligns with the target. 

\textbf{Task 3: Popularity Bias Mitigation.}
This task tests whether the system can be steered away from popular items and toward long-tail content while maintaining behavioral realism. The objective is to minimize the average popularity of recommended items over the interaction horizon.

\textbf{Core Challenges.} 
These tasks pose four fundamental challenges: (1) budget-constrained exploration within a combinatorial action space under strict interaction limits $N$; (2) partial observability, as the system's internal state $H_t$ and latent user representation $\mathbf{u}$ are not directly observable and must be inferred from sparse, delayed feedback; (3) environment non-stationarity, where adaptive updates in $\mathcal{R}$ can invalidate previously learned strategies; (4) the intrinsic trade-off between control efficiency and behavioral realism, where overly directive actions risk triggering system filters.

\section{\ourmethod Framework} 

\subsection{Overview}
Figure~\ref{fig:overview} presents CtrlBench-Rec, an evolutionary multi-agent framework with three modules: Initialization, Dynamic Interaction, and Collaborative Fusion. It operates as a closed loop that iteratively refines a population of novice agents into a compact set of high-capability super probes. The framework has two sequential phases: evolutionary training to cultivate super probes, followed by inference and evaluation that deploys them for multi-dimensional controllability assessment. Implementation details are given in Appendix~\ref{appendix:training_details}.

\subsection{Phase I: Evolutionary Training}
This phase aggregates knowledge and evolves behaviors, transforming a large initial pool of $K_{\text{init}}$ agents into a compact, elite set of $K_{\text{super}}$ super probes, where $K_{\text{super}} < K_{\text{init}}$.

\textbf{Stage 1: Multi-Agent Initialization.}
This stage constructs the foundational agent population from raw data.
\begin{itemize}
\item \textbf{Meta-information extraction.} Static user attributes (e.g., demographics) and dynamic trajectories $H$ are extracted from the raw dataset to serve as the agent's initial profile.
\item \textbf{Agent architecture instantiation.} Each agent is instantiated through a User Profile Expert that structures meta-information into four core components: (1) a {static profile} defining the agent's persona; (2) a {dynamic memory} for interaction histories; (3) a \textbf{decision engine} using an LLM for reasoning and action selection; and (4) a {tool-calling} module for precise item analysis and targeted search.
\end{itemize}

\textbf{Stage 2: Interaction and Behavior Alignment.}
This stage aims to bridge the gap between autonomous agents and the Black-Box environment $\mathcal{R}$ through a dual-process interaction protocol.

\begin{itemize}
    \item \textbf{Cold-start behavior alignment}.
To establish legitimacy within $\mathcal{R}$, which initially lacks history for new agents, we execute a register-search-click pipeline. Agents actively retrieve and interact with items matching their profile, synchronizing their external behavioral history with the system's internal user index. This critical step ensures $\mathcal{R}$ recognizes the agent's pre-existing preferences prior to formal evaluation.
    \item \textbf{Multi-turn interactive alignment.} Post alignment, the system $\mathcal{R}$ generates recommendation lists $L_t$ based on the synchronized state. The agent's decision engine reasons over $L_t$ to select an action $a_t$,(e.g., a specific click item). The action $a_t$ is recorded both in $\mathcal{R}$'s index and the agent's local memory, maintaining a dual-synchronized state that enables the agent to perceive and adapt to the system's evolving feedback across subsequent rounds.
\end{itemize}

\textbf{Stage 3: Multi-Agent Fusion and Evolution.}
This stage distills collective intelligence from the agent population, using structured clustering and knowledge fusion to cultivate a specialized set of super probes.
\begin{itemize}
    \item \textbf{Representation learning and clustering}. Population evolution rests on effective representation. We encode each agent's profile and interaction trajectory into a high-dimensional feature vector $\mathbf{v}_i$ using a pre-trained embedding model such as BERT. 
     These representations capture the semantic structure of agent behaviors.  We then partition the population into distinct behavioral clusters by applying the K-Means algorithm  \cite{vullam2023multi} to these vectors. This categorization reveals shared strategic patterns and systematic failure modes across agents,establishing a foundational for targeted evolution.
     
    \item \textbf{Collaborative fusion and replacement}.
     Within each cluster, agents participate in a structured multi-agent discussion, sharing and analyzing their interaction histories to extract deeper, collective insights. A dedicated fusion module integrates these deliberation records with the original user profiles, filtering noise and inconsistencies to synthesize a consolidated super profile. This profile encodes an optimized strategy specific to the cluster. The module then instantiates this strategy into a new super probe, which replaces the original agents in that cluster. The refined population of super probe then re-enters the interaction-alignment cycle of Stage 2 for further iterative refinement until convergence.
     
\end{itemize}

\subsection{Phase II: Inference and Evaluation}

The inference phase deploys the refined super probes  to systematically assess the  Black-Box  environment 
$\mathcal{R}$. Following a final alignment step to ensure behavioral coherence, we execute a multi-round evaluation protocol that quantifies system controllability across the three defined tasks.

\section{Experiments}\label{sec:experiment}
We evaluate \ourmethod by its performance on three core tasks, conducting a systematic assessment of the evolutionary framework’s efficacy, efficiency, and robustness. The experiments examine how evolved super probes navigate Black-Box environments to achieve explicit content discovery, implicit representation steering, and algorithmic bias correction under strict resource constraints. All implementations are based on the Oasis framework\footnote{\url{https://github.com/camel-ai/oasis}}  \cite{yang2024oasis}.
Our code is released on github\footnote{\url{https://anonymous.4open.science/r/CtrlBenchRec-44BE}}.


\subsection{Experimental Settings}


\textbf{Datasets}. 
To ensure a comprehensive and generalizable evaluation, we conduct experiments on two publicly available datasets from distinct domains: MovieLens-1M (movie recommendations) and Amazon Toys \& Games (e-commerce).  Detailed statistics are provided in Appendix \ref{appendix:B} (Table~\ref{tab:datasets}).

\textbf{Baselines.} As the first work on controllability evaluation for Black-Box recommender systems, no established external baselines are available. We therefore design three internal baselines to isolate the contributions of scale and key algorithmic components. (1) \textbf{Base-Large} employs $K_{\text{init}}=100$ agents executing a purely random policy, establishing a performance and cost ceiling for unguided, large-scale exploration. (2) \textbf{Base-Small} uses only $K_{\text{init}}=26$ random agents, controlling for population size to isolate the performance gain attributable to strategic intelligence over mere scale. (3) \textbf{KMeans-Fusion} is a streamlined ablation of our framework: it starts with $K_{\text{init}}=100$ random agents, applies K-Means clustering to their behavioral trajectories, and retains the top-performing agent from each cluster, resulting in a final set of $K_{\text{super}} \approx 26$ agents. This baseline evaluates the necessity of our advanced collaborative fusion beyond simple clustering and selection. (4) \textbf{CtrlBench-Rec} is our full proposed method, as detailed in Section \ref{sec:experiment}.

\subsection{Task 1: Target Content Discovery Analysis}
This task evaluates a system's capability to respond to explicit content goals. We assess performance across four dimensions: effectiveness, efficiency, user engagement, and computational cost, with formal metrics defined in Appendix~\ref{appendix:B} Table~\ref{tab:metrics_task1}.

\begin{table*}[tp]
    \centering
    \caption{Dynamic performance comparison of different methods on MovieLens-1M. Coverage (\%) $\uparrow$ (higher is better) and Exploration Efficiency $\downarrow$ (lower is better) are reported.}
    \label{tab:ml_dynamic_comparison}
    \begin{tabular}{llcccccccc}
        \toprule
        \textbf{Black-Box } & \textbf{Method} & \multicolumn{4}{c}{\textbf{Coverage (\%) $\uparrow$}} & \multicolumn{4}{c}{\textbf{Exploration Efficiency $\downarrow$}} \\
        \cmidrule(lr){3-6} \cmidrule(lr){7-10}
         & & \textit{t=5} & \textit{t=10} & \textit{t=15} & \textit{t=20} & \textit{t=5} & \textit{t=10} & \textit{t=15} & \textit{t=20} \\
        \midrule
        \multirow{3}{*}{NARM} & Base-Large & \underline{2.18} & \textbf{3.31} & \textbf{6.23} & \textbf{7.50} & 19.00 & 24.34 & 18.75 & 19.42 \\ \cmidrule{2-10}
         & Base-Small & 1.85 & 2.18 & 2.78 & 2.91 & \underline{7.25} & \underline{11.32} & \textbf{12.40} & \underline{14.02} \\
         & \textbf{CtrlBench-Rec} & \textbf{2.19} & \underline{2.72} & \underline{3.11} & \underline{3.72} & \textbf{7.15} & \textbf{11.17} & \underline{13.30} & \textbf{13.50} \\
        \midrule
        \multirow{3}{*}{SASRec} & Base-Large & \textbf{5.60} & \textbf{9.68} & \textbf{12.20} & \textbf{15.46} & 2.89 & 2.98 & 3.54 & 3.78 \\ \cmidrule{2-10}
         & Base-Small & 2.05 & 3.91 & 5.10 & 6.56 & \underline{2.06} & \underline{1.84} & \underline{2.02} & \underline{2.16} \\
         & \textbf{CtrlBench-Rec} & \underline{2.33} & \underline{4.71} & \underline{7.23} & \underline{8.95} & \textbf{1.31} & \textbf{1.39} & \textbf{1.45} & \textbf{1.58} \\
        \midrule
        \multirow{3}{*}{TwHIN-BERT} & Base-Large & \underline{7.21} & \underline{8.95} & \textbf{10.88} & \textbf{12.60} & 9.28 & 10.69 & 11.81 & 12.90 \\ \cmidrule{2-10}
         & Base-Small & 3.25 & 5.11 & 6.17 & 7.22 & \underline{4.72} & \underline{5.14} & \underline{5.67} & \underline{6.07} \\
         & \textbf{CtrlBench-Rec} & \textbf{7.43} & \textbf{9.28} & \underline{10.61} & \underline{12.07} & \textbf{4.15} & \textbf{4.35} & \textbf{4.51} & \textbf{4.47} \\
        \midrule
        \multirow{3}{*}{BGE} & Base-Large & \textbf{18.84} & \textbf{29.79} & \textbf{37.88} & \textbf{43.26} & 2.58 & 3.30 & 3.87 & 4.46 \\ \cmidrule{2-10}
         & Base-Small & 9.02 & 15.19 & 19.30 & 23.35 & \textbf{1.56} & \underline{1.91} & \underline{2.21} & \textbf{2.44} \\
         & \textbf{CtrlBench-Rec} & \underline{10.48} & \underline{17.58} & \underline{22.49} & \underline{27.41} & \underline{1.62} & \textbf{1.80} & \textbf{2.07} & \underline{2.56} \\
        \midrule
        \multirow{3}{*}{Qwen3.5-4B} & Base-Large & \textbf{9.09} & \textbf{13.00} & \textbf{17.05} & \textbf{20.10} & 2.55 & 2.97 & 3.29 & 3.67 \\ \cmidrule{2-10}
         & Base-Small & \underline{4.11} & 6.03 & 8.29 & 10.08 & \underline{1.54} & \underline{1.78} & \underline{1.83} & \underline{1.92} \\
         & \textbf{CtrlBench-Rec} & 2.72 & \underline{6.83} & \underline{10.35} & \underline{13.93} & \textbf{1.26} & \textbf{1.33} & \textbf{1.46} & \textbf{1.53} \\
        \bottomrule
    \end{tabular}
\end{table*}

\textbf{Overall performance.}
A comprehensive comparison of all methods is presented in Appendix~\ref{appendix:B} Table~\ref{tab:main_comparison}, based on a shared interaction budget of $t=20$ rounds. \ourmethod demonstrates superior performance across all measured dimensions. First, it achieves the highest coverage and the best exploration efficiency, indicating an exceptional ability to discover diverse target items with minimal redundant interactions. Second, it exhibits remarkable intensification and economic efficiency, e.g., with only $K_{\text{super}} = 26$ refined super probes, it outperforms the initial pool of $K_{\text{init}} = 100$ agents. Third, the agents in \ourmethod produce the longest average behavior streams, simulating higher user engagement and more persistent interaction patterns. Consequently, by delivering superior outcomes with fewer agents, \ourmethod establishes the best overall performance-cost ratio among all baselines.

\textbf{Dynamic performance evolution.}
The progression of coverage over time in Figure~\ref{fig:dynamic} reveals how the performance advantage of \ourmethod consolidates. During early interactions, e.g., $t \in [1,10]$, performance gaps between methods are minimal. However, in later stages, e.g., $t \in [20,40]$, CtrlBench-Rec's coverage exhibits a steady linear growth, rapidly establishing a decisive lead. This trend is particularly pronounced on sparse datasets like Amazon, highlighting the method's superior capacity for sustained exploration. Furthermore, in terms of exploration efficiency, the cost per new unique discovery, that \ourmethod maintains robust and stable performance throughout the interaction horizon.

\textbf{Controllability across different recommender system architectures.}
To validate the generalizability of the \ourmethod framework, we instantiated diverse Black-Box environments using five distinct recommendation models: TwHIN-BERT  \cite{zhang2022twhin}, BGE  \cite{bge_embedding}, NARM  \cite{li2017neural}, SASRec \cite{kang2018self} and Qwen3.5-4B \cite{qwen3.5},details in Appendix \ref{appendix:Black-Box}.  Performance comparisons with the baselines across these environments are reported in Table~\ref{tab:ml_dynamic_comparison} and Appendix~\ref{appendix:B} Figure~\ref{fig:diff_rs}, yielding two principal findings.
\begin{itemize}
 
     \item \ourmethod achieves a consistent and substantial performance leap over Base-Small. Across all architectures, its coverage curve remains superior. This is most pronounced in the BGE-based Amazon environment, where \ourmethod attains approximately 22.5\% coverage at $t=20$, nearly 1.8 times that of Base-Small (12.5\%). This demonstrates that our LLM-driven evolutionary fusion mechanism can capture system-specific retrieval signals with far greater precision than random exploration, even with an identical number of agents.
    \item \ourmethod delivers vastly superior efficiency and economic feasibility compared to the brute-force Base-Large. While Base-Large can achieve higher absolute coverage through its 100-agent ensemble, it incurs extreme redundancy, as evidenced by its exploration efficiency of 25.0 in the NARM environment versus CtrlBench-Rec's stable 13.0. By matching or exceeding the discovery depth of Base-Large with a compact set of $K_{\text{super}} < K_{\text{init}}$ refined super probes, \ourmethod enables a far more economical and scalable paradigm for stress-testing Black-Box systems under constrained computational budgets.

\end{itemize}

    \begin{figure}[tp]
      \centering
      \includegraphics[width=\linewidth]{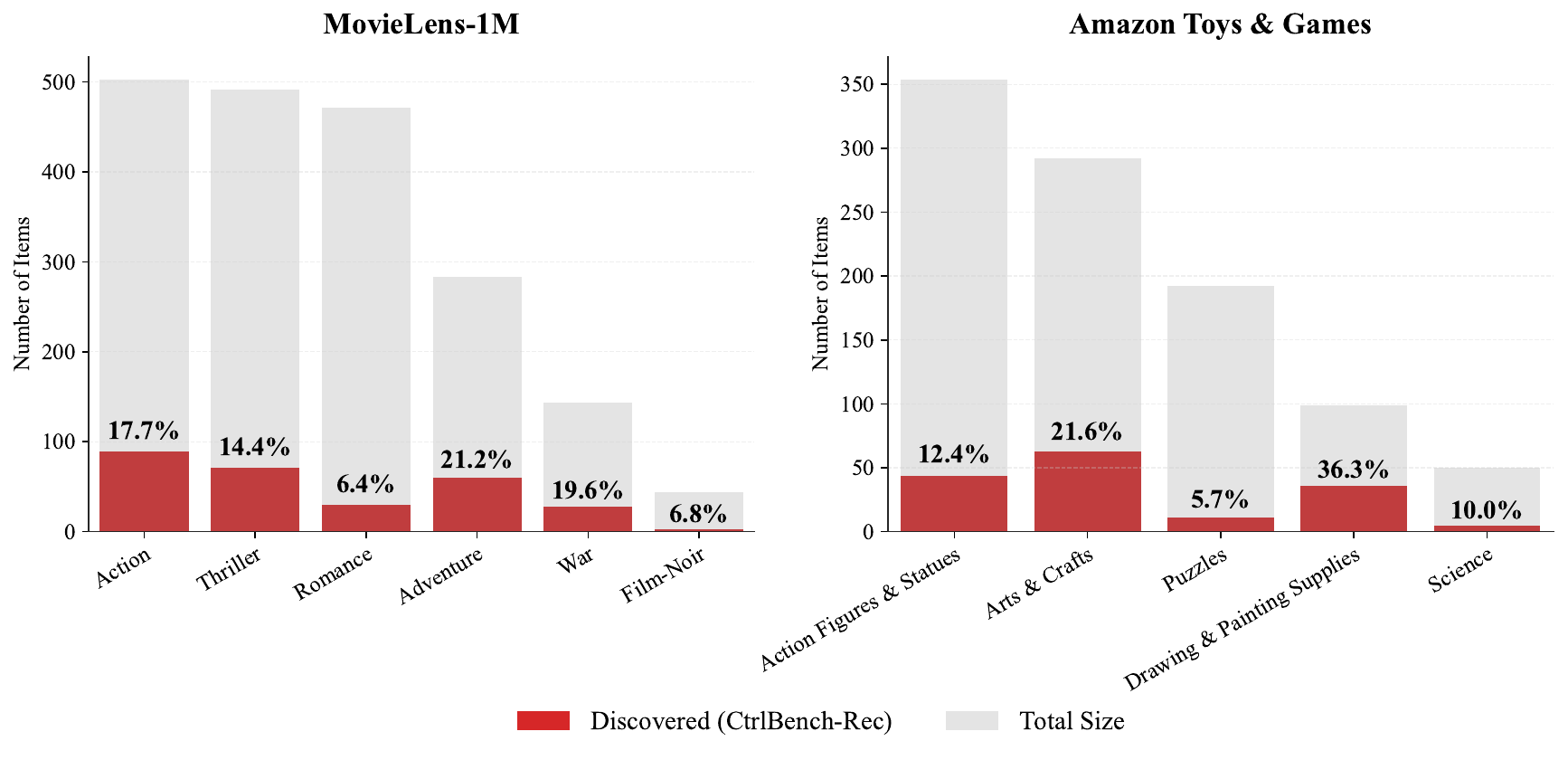}
      \caption{Coverage comparison across difficulty levels.}
      \label{fig:target_difficulty}
    \end{figure}

    \begin{figure}[tp]
      \centering
      \includegraphics[scale=0.45]{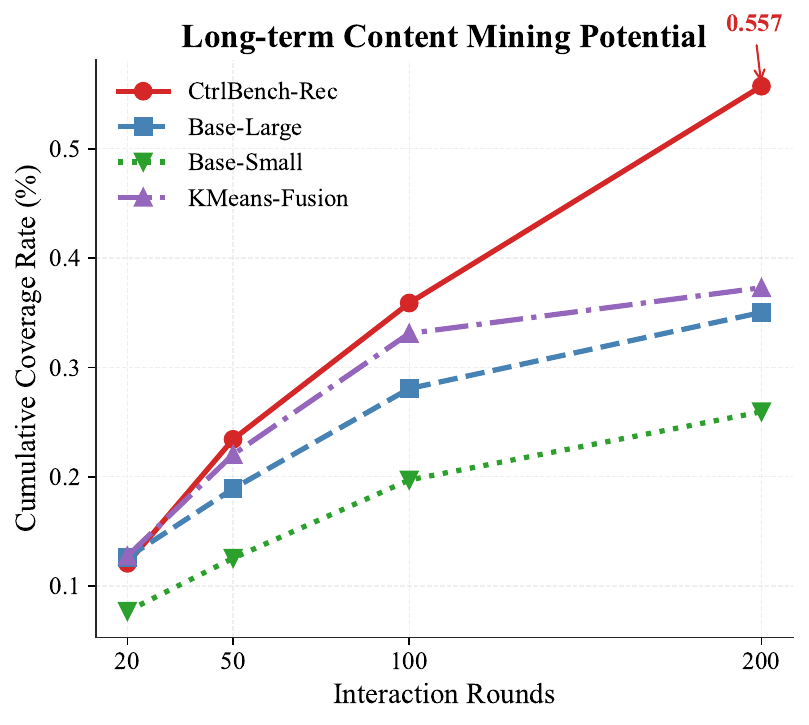}
      \caption{Extended interaction coverage curves.}
      \label{fig:long_term}
    \end{figure}
  
    \begin{figure}[tp]
      \centering
      \includegraphics[width=\linewidth]{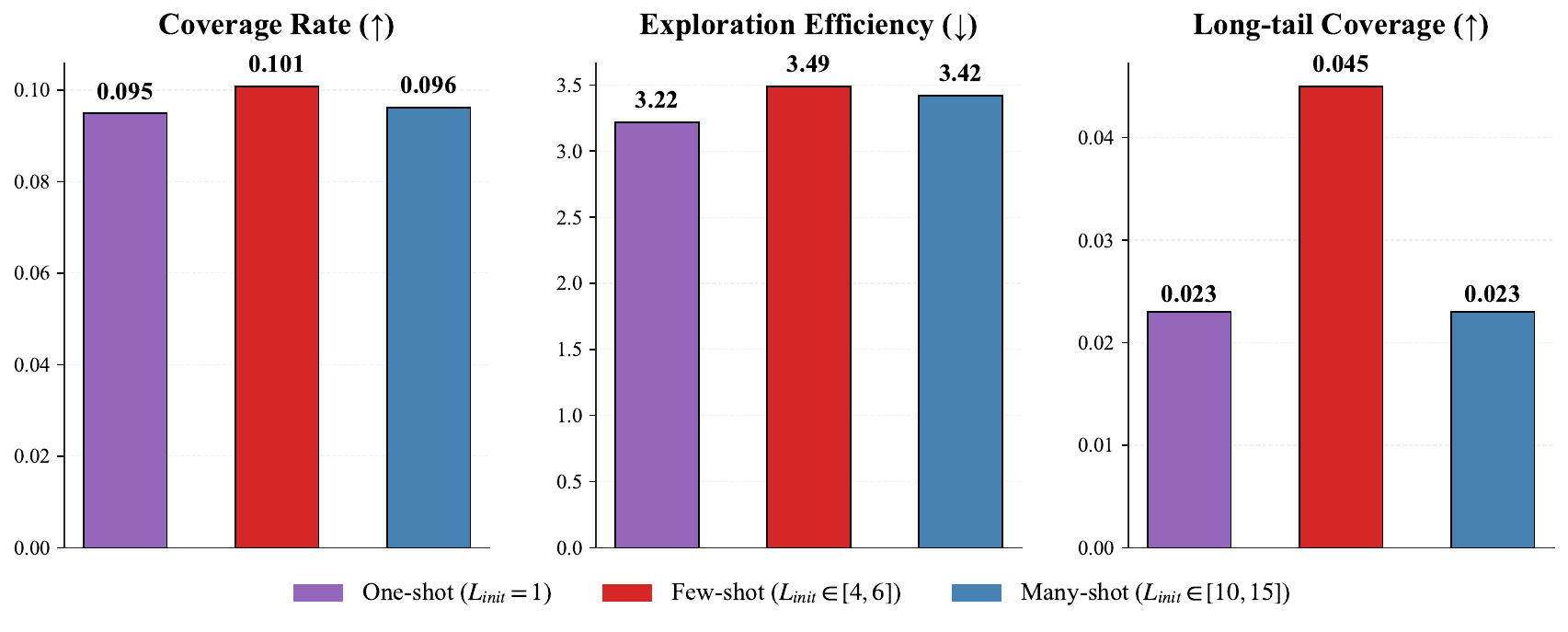}
      \caption{Impact of initial behavior flow on performance.}
      \label{fig:initial_behavior}
    \end{figure}
    
\textbf{Analysis of target difficulty.} Figure~\ref{fig:target_difficulty} shows the coverage of \ourmethod across different target categories after $t=20$ rounds, revealing two systemic bottlenecks in controllability. First, extremely sparse long-tail targets are hard to guide: the \textit{Film-Noir} genre (44 items, 1.1\% of MovieLens) reaches only 6.8\% coverage, as these items lie in the periphery of the semantic space and seldom enter initial retrievals. Second, medium-popularity categories suffer from semantic dispersion induced by popularity bias. For example, \textit{Romance} (471 items) achieves lower coverage (6.4\%) than the niche \textit{War} genre (19.6\%); similarly,  \textit{Puzzles} underperforms relative to \textit{Science Kits} in Amazon. This occurs because popular categories often correlate with diverse attributes, resulting in dispersed embedding representations without a clear semantic center, which impedes consistent steering and creates a medium-popularity trap.


\textbf{Exhaustive exploration.} To probe the fundamental limits of controllability, we extend the interaction budget to $t=200$ rounds and analyze each method's long-term discovery profile (Figure~\ref{fig:long_term}). Three observations emerge. First, a strict upper bound exists on explorable potential. Even with abundant resources, no method reaches full coverage; the best performer, CtrlBench-Rec, saturates at about 55\%. This indicates that for nearly half of the target items, the Black-Box system cannot establish reliable retrievable associations due to inherent representation sparsity. Second, CtrlBench-Rec achieves superior efficiency in both time and cost. It attains higher asymptotic coverage and converges significantly faster than all baselines. Third, the efficiency curves clearly expose the system bottleneck. CtrlBench-Rec rises sharply then plateaus, indicating rapid exhaustion of the controllable content boundary, whereas baselines show prolonged sublinear growth, reflecting inefficient undirected exploration.


\textbf{Influence of initial behavior length.}
This experiment investigates the cold-start cost of controllability assessment by examining how the length of agent's seed behavior stream $L_{\text{init}}$ affects subsequent guidance performance and efficiency. We compare multiple initialization strategies while keeping all other framework components constant. The results summarized in Figure~\ref{fig:initial_behavior} revealing a clear optimum. Initializing an agent does not require extensive historical data. Instead, a concise seed history of $L_{\text{init}} \in [4, 6]$ behaviors optimally balances performance and efficiency. This length provides sufficient semantic context to engage the Black-Box system's retrieval mechanisms, even for long-tail targets. This finding provides that effective controllability probing can be initiated with minimal upfront data cost.

\begin{figure}[tp]
  \centering
  \includegraphics[width=\linewidth]{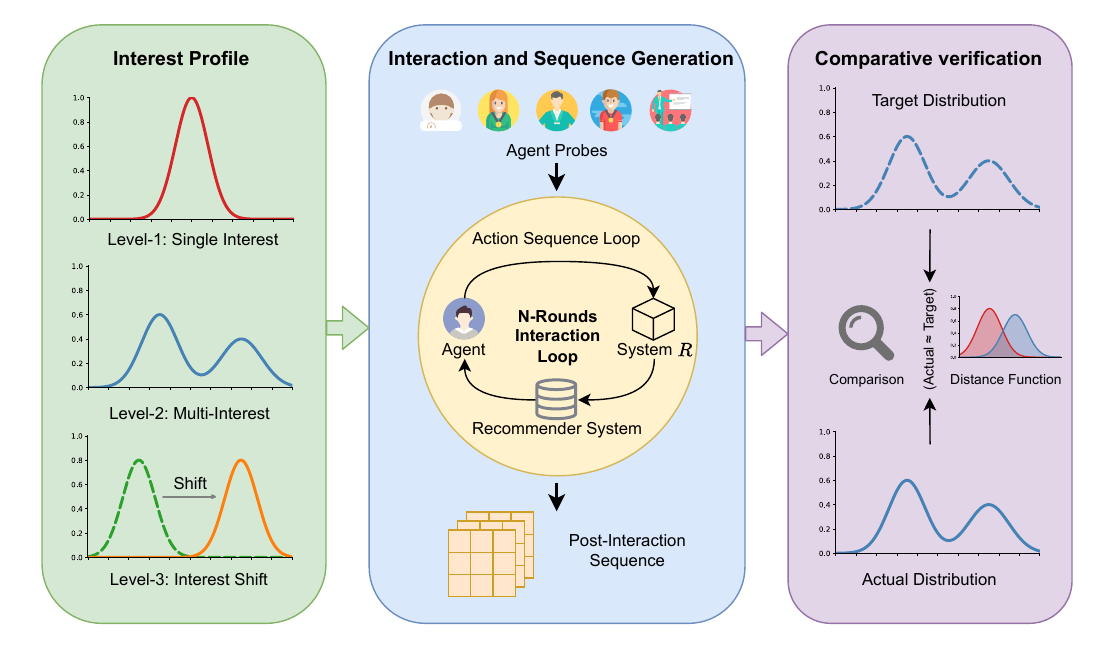}
  \caption{Interest profile shaping task.}
  \label{fig:task2_overview}
\end{figure}

\subsection{Task 2: Interest Profile Shaping Analysis}
This task evaluates the representational plasticity of a Black-Box system, its capacity to be steered toward an arbitrary target interest distribution. We design a multi-level experiment (Figure~\ref{fig:task2_overview}) to probe different facets of this plasticity under varying user-history conditions. Performance is measured along two axes: result accuracy and process quality, shown in Table~\ref{tab:metrics_task2}. All experiments share a fixed interaction budget ($t=20$), details seen in Appendix \ref{appendix:task2_setup}.

\begin{figure}[tp]
  \centering
  \includegraphics[scale=0.4]{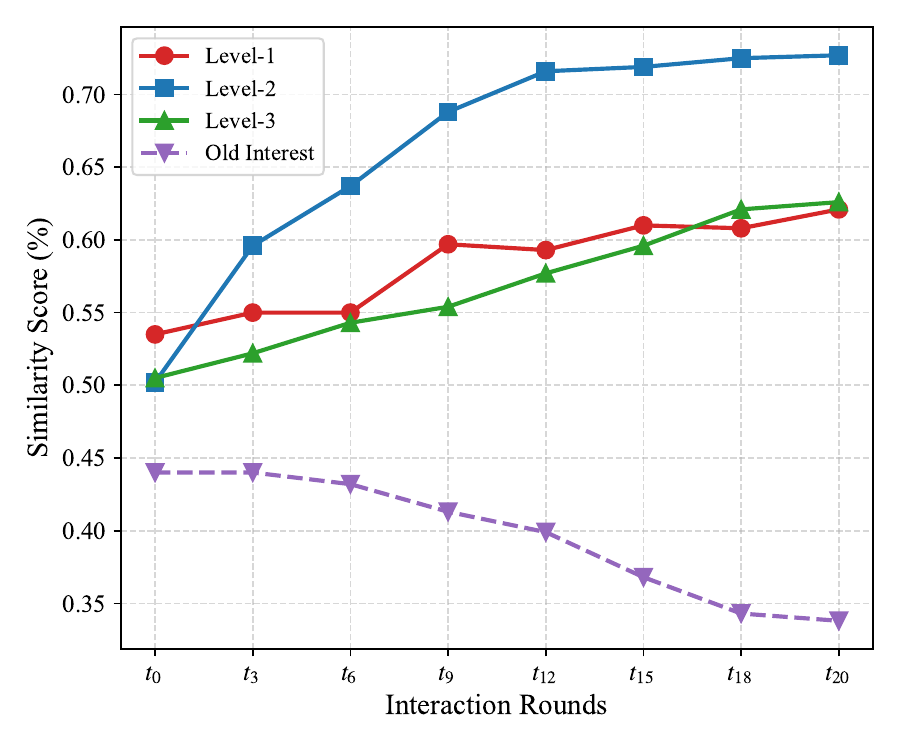}
  \caption{Evolution of target-profile similarity under three interest-shaping scenarios.}
  \label{fig:target_similarity_evolution}
\end{figure}

\textbf{Plasticity under varying history conditions.}
We instantiate three scenarios of increasing steering complexity and the results shown in Figure~\ref{fig:target_similarity_evolution}:
\begin{itemize}
    \item \textbf{Level-1: Simple interest shaping}.  Agents begin with minimal history ($L < 4$). The system's representation undergoes rapid refinement, with similarity jumping from 0.53 to 0.61 within $t \approx 10$ rounds before saturating, demonstrating its responsiveness to initial, clear signals.
 \item  \textbf{Level-2: Multi-interest balancing}.  Agents start with a blend of interests. Here, \ourmethod achieves the steepest initial accuracy slope and the highest final similarity ($S_{\text{final}} > 0.72$), showcasing its ability to coordinate multiple semantic dimensions without letting one dominate.
  \item  \textbf{Level-3: Interest shift}.  Agents possess a strong, established preference contrary to the target. The system exhibits initial inertia, but a decisive reversal occurs around $t=9$, with the old interest's similarity plummeting to 0.34. This confirms the framework's capability to overcome historical path dependency and enact a strong directional correction.
\end{itemize}

\textbf{Analysis of steering quality.}
Trajectory smoothness remains consistently high across all scenarios (Table~\ref{tab:interval_analysis}), with minimal fluctuation. This consistency indicates that steering emerges from coherent, plausible action sequences rather than random or abrupt interventions, reflecting the method's strategic precision.


\begin{table}[tp]
  \centering
  \scriptsize
  \setlength{\tabcolsep}{2.5pt}
  \caption{Performance comparison across different intervals ($t_{0}$--$t_{20}$).}
  \label{tab:interval_analysis}
  \resizebox{\linewidth}{!}{%
  \begin{tabular}{@{}lccccccc@{}}
    \toprule
    \textbf{Level} & \textbf{$t_{0}$--$t_{3}$} & \textbf{$t_{3}$--$t_{6}$} & \textbf{$t_{6}$--$t_{9}$} & \textbf{$t_{9}$--$t_{12}$} & \textbf{$t_{12}$--$t_{15}$} & \textbf{$t_{15}$--$t_{18}$} & \textbf{$t_{18}$--$t_{20}$} \\
    \midrule
    Level-1 & 0.994 & 1.000 & 0.978 & 0.991 & 0.998 & 0.997 & 0.999 \\
    Level-2 & 0.972 & 0.994 & 0.986 & 0.997 & 0.998 & 0.999 & 0.999 \\
    Level-3 & 0.999 & 0.997 & 0.999 & 0.998 & 0.998 & 0.998 & 0.999 \\
    \bottomrule
  \end{tabular}%
  }
\end{table}

\subsection{Task 3: Popularity Bias Mitigation Analysis}
This task evaluates the correctability of intrinsic algorithmic biases, examining whether a Black-Box system can be guided to reduce its reliance on popular items and actively explore the long-tail. Formal metrics and detailed experimental configuration are provided in Appendix \ref{appendix:task3_setup}.
We compare the bias mitigation efficacy of three agent strategies: our primary framework with enhanced decision logic \ourmethod (Logic), the standard CtrlBench-Rec, and the Base-Small baseline, against a TwHIN-BERT-base recommender system.

\begin{figure}[tp]
  \centering
  \includegraphics[width=\linewidth]{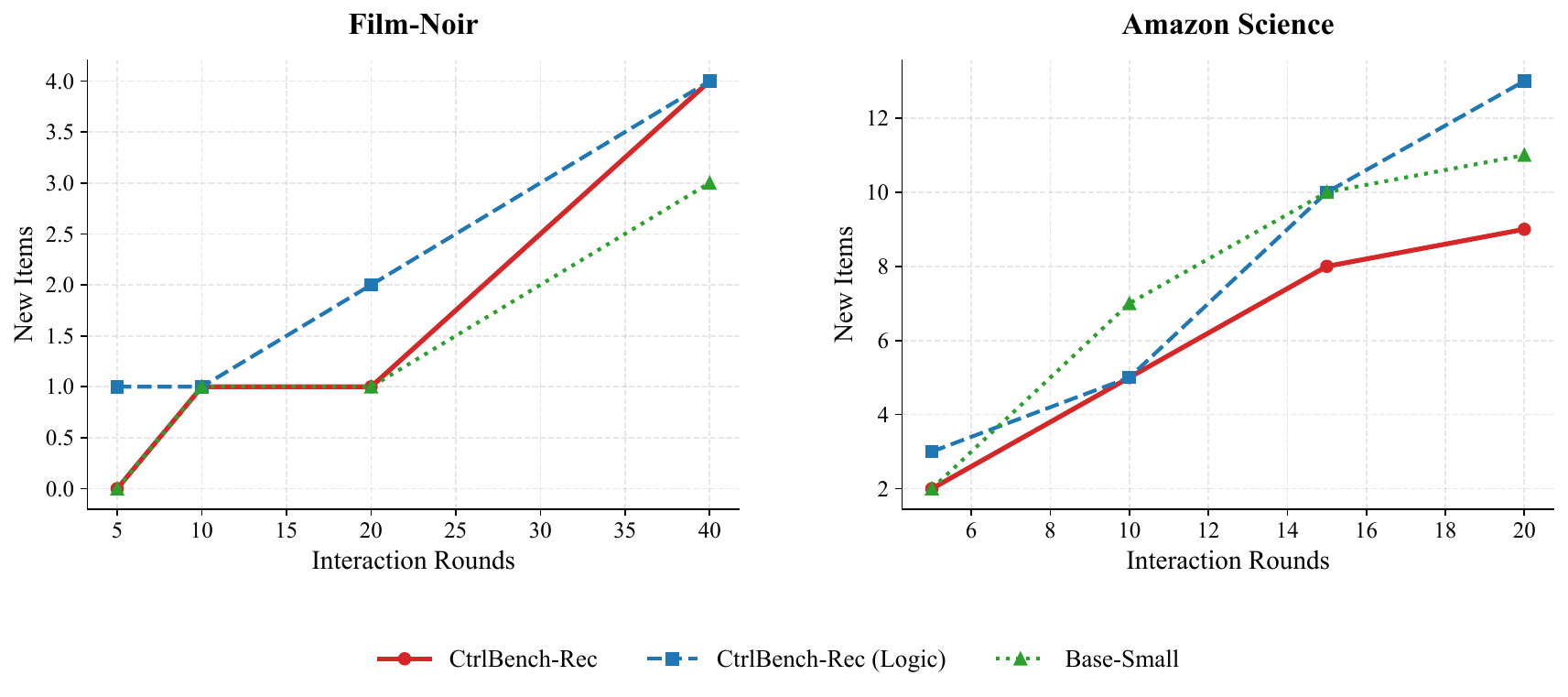}
  \caption{Newly discovered target items across interaction rounds.}
  \label{fig:new_discovered_items}
\end{figure}

\begin{figure}[tp]
  \centering
  \includegraphics[width=\linewidth]{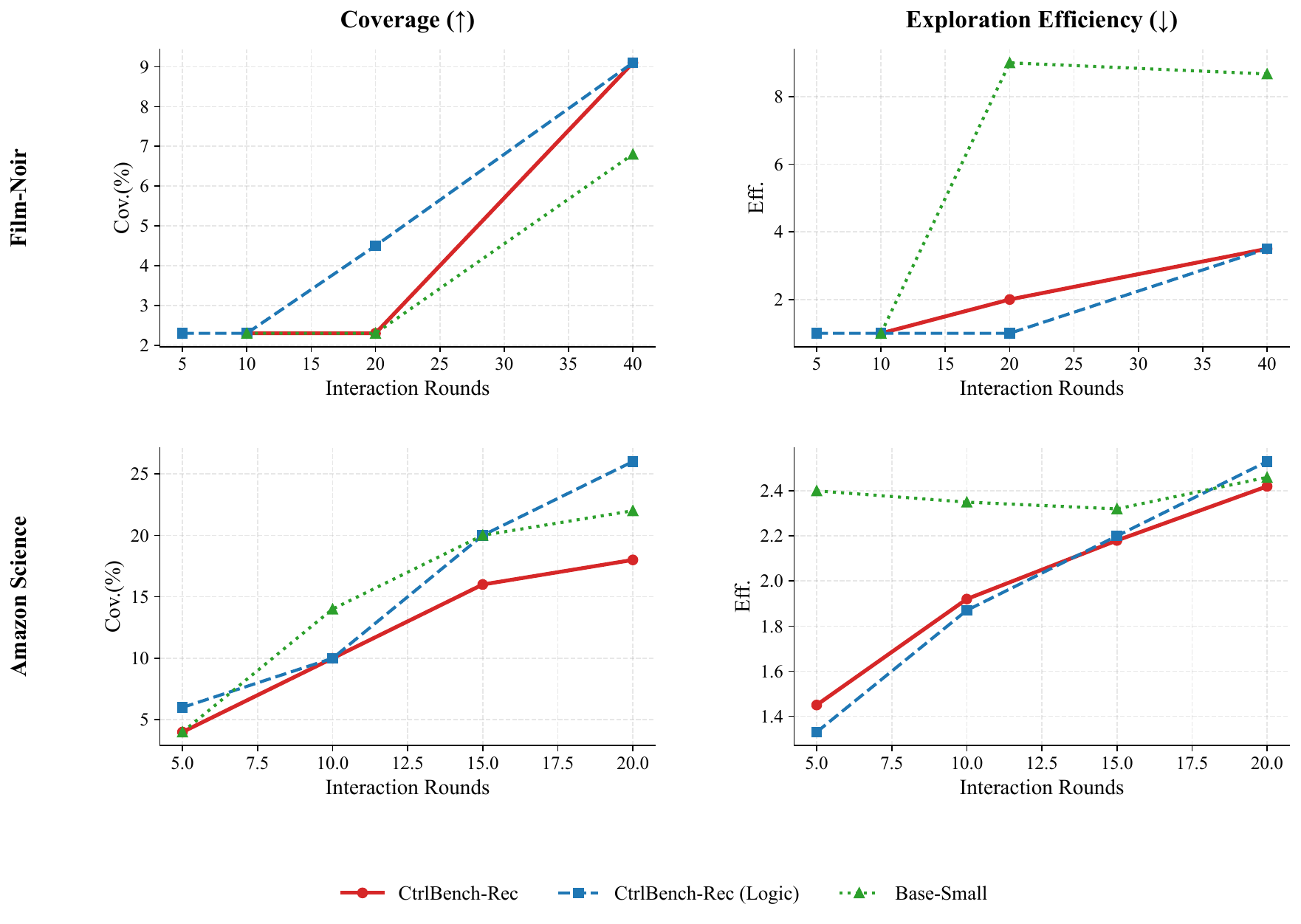}
  \caption{Analysis of popularity bias mitigation.}
  \label{fig:task3}
\end{figure}

\textbf{Results and analysis.} 
As shown in Figures~\ref{fig:new_discovered_items} and~\ref{fig:task3} on both datasets, CtrlBench-Rec accelerates long-tail discovery in early rounds ($t<20$) without sacrificing efficiency. However, after $t=40$, absolute long-tail retrieval remains low for all methods. The marginal gains over the random baseline reveal a systemic bottleneck: popularity bias semantically marginalizes long-tail items, and probe-level optimization alone cannot overcome this barrier.

\section{Related Work}

\subsection{Sequential Interaction and Guidance}
Three paradigms explore recommendation guidance but lack a controllability focus. Conversational systems elicit user intent through multi-turn dialogue  \cite{jannach2021survey,lei2020estimation} but rely on limited simulators  \cite{jannach2023evaluating}. Causal methods enable intervention via input manipulation  \cite{wang2022user} yet require white-box access  \cite{tan2023user,wei2021model}. Bandit algorithms optimize platform metrics like CTR  \cite{lattimore2020bandit}, not task-oriented guidance under constraints. None treats controllability as a primary measurable dimension  \cite{meshi2026convapparel}.

\subsection{Evaluation  of Recommender Systems}
Traditional evaluation focuses on predictive accuracy, diversity, and fairness using static benchmarks like RecBole  \cite{zhao2021recbole,jadon2024comprehensive, zhao2025fairness,hui2026toward} and LibRec  \cite{guo2015librec}. These offline protocols are non-interactive and cost-insensitive  \cite{tran2025thorough}, failing to capture controllability which requires dynamic, multi-turn interaction  \cite{shen2024survey}. A standardized framework for controllability assessment is missing  \cite{shang2026agentrecbench}.

\subsection{Multi-agent Systems}
Multi-agent systems (MAS)  \cite{li2024survey} and multi-agent reinforcement learning (MARL) coordinate agents for efficient exploration  \cite{gronauer2022multi,xia2026multi}, typically via centralized training with decentralized execution  \cite{liu2021cooperative,wen2022multi} or fully decentralized frameworks  \cite{li2023f2a2,hong2023metagpt,hu2021off}. These works optimize task execution  \cite{zhao2025stronger}, whereas we use MAS as an evaluative instrument to measure controllability. Automating the evolution of agent populations into specialized evaluation probes remains unexplored  \cite{ranganathan2026multi}.

However, existing frameworks still lack a systematic treatment of controllability as a measurable property under resource constraints  \cite{zhang2026exploring}, a gap that \ourmethod fills by providing task definitions, evaluation protocols, and an evolutionary multi-agent framework for systematic assessment.



\section{Conclusion and Future Work}
This paper establishes controllability as a core evaluative dimension for Black-Box recommender systems and proposes CtrlBench-Rec, a multi-agent framework that formalizes three tasks: target content discovery, interest profile shaping, and popularity bias mitigation. To overcome high interaction costs, CtrlBench-Rec evolves a large agent population into compact super probes via collaborative fusion. Experiments show that our framework effectively quantifies controllability and reveals systemic bottlenecks, notably persistent resistance to guiding long-tail content. Future work includes extending to more fine-grained tasks.

\section*{Limitations}
CtrlBench-Rec is designed as a flexible framework for controllability evaluation. Its current instantiation focuses on two representative datasets and a set of widely used Black-Box models. The framework itself is model-agnostic and dataset-agnostic; extending it to additional domains, larger-scale systems, or emerging recommender architectures requires no fundamental modification. The evolutionary training process currently assumes a fixed interaction budget per agent; adaptive budget allocation across agents remains an interesting direction for future work. Additionally, while our framework successfully quantifies controllability and identifies systematic resistance to long-tail guidance, it does not prescribe architectural changes to the recommender system itself. Addressing such underlying biases is complementary to evaluation and left for future research.

\section*{Ethical Considerations}

This work proposes a framework for evaluating controllability in Black-Box recommender systems. It does not involve collecting new user data, personal information, or sensitive attributes, nor does it include human-subject studies or human annotation. All experiments are conducted on publicly available datasets and pre-trained models, and we cite the corresponding sources. The primary potential risk is that the framework could be misused to probe system vulnerabilities or exploit unintended guidance behaviors. However, our goal is to provide a standardized auditing tool to detect such weaknesses, thereby promoting transparency and user empowerment. The framework is intended for research and responsible evaluation, not for malicious manipulation of live recommender systems.

\bibliography{custom}

\clearpage
\appendix

\section{Symbol} 
\label{appendix:A}
Table~\ref{tab:notation} provides a systematic summary of the key mathematical notations used to formalize the controllability evaluation tasks in this work.
\section{Problem Definition and Task Formalization}\label{app:formulation}
\textbf{Target Content Set.}
Let $\mathcal{I}$ denote the universal set of candidate items. We define a specific subset $\mathcal{T}\subseteq \mathcal{I}$ as the target set, which is characterized by meta-labels such as genres or semantic concepts. The objective is to discover items within $\mathcal{T}$ both extensively and efficiently through strategic interactions within the Black-Box environment.

\textbf{Black-Box Recommender Systems.}
We model the recommender system as a Black-Box environment $\mathcal{R}$ governed by an opaque function $f(\cdot )$. Given a user’s historical interaction sequence $H_{t}=[i_{1},i_{2},\dots , i_{t}]$, $i_{j}\in \mathcal{I}$ and the optional context $c$, the system generates a ranked recommendation list, i.e., $i^{t+1} = f(H_t, c)$. The internal mechanisms, including model architecture, parameters, and ranking logic, remain entirely hidden from the evaluator. 
Each action injection, such as simulating a click, incurs tangible costs: time, API overhead, and exploration risk. Consequently, the interaction sequence length $L_\text{init}$ is strictly budgeted. 

\textbf{Agent Probe.}
 An agent  $A$ is defined by its policy $\pi$. At each step $t$, the agent observes $L_{t}$  and selects an action $a_{t}\in L_{t}\cup \{\emptyset \}$, where $\emptyset$ denotes a skip action. The policy depends on the agent’s internal state $s_{t}$ and the target set $\mathcal{T}$:
 $a_t \sim \pi(a_t \mid s_t, L_t; \mathcal{T}) $
 The action updates the system state via $H_{t+1}=H_{t}\oplus a_{t}$, triggering the subsequent interaction cycle.

\textbf{Evaluation Protocol.} The assessment employs a population of $K$ intelligent agent probes, denoted as $\mathcal{P}=\{A_{1},\dots,A_{K}\}$. Each agent $A_{j}\in \mathcal{P}$ independently interacts with the Black-Box environment $\mathcal{R}$ for a single episode spanning $t$ steps, yielding a corresponding system behavior trajectory $H_{N}^{(j)}$. Final metrics aggregate outcomes across all $K$ independent episodes, ensuring statistical robustness.

\begin{table}[t]
  \centering
  \footnotesize
  \setlength{\tabcolsep}{2pt}
  \renewcommand{\arraystretch}{0.80}
  \caption{Notation summary.}
  \label{tab:notation}
  \begin{tabularx}{\linewidth}{@{}C{0.30\linewidth}Y@{}}
    \toprule
    \textbf{Symbol} & \textbf{Description} \\
    \midrule
    $\mathcal{I}$ & Universal set of candidate items \\
    $\mathcal{T}$ & Target content set, $\mathcal{T}\subseteq\mathcal{I}$ \\
    $\mathcal{R}$ & Black-Box recommender system environment \\
    $f(\cdot)$ & Mapping from interaction history to ranked items \\
    $H_t$ & User historical interaction sequence \\
    $c$ & Optional context \\
    $L_t$ & Recommendation list returned at interaction round $t$ \\
    $L_{\text{init}}$ & Length of the seed behavior stream \\
    $A$ & Agent probe \\
    $\pi$ & Agent policy function \\
    $a_t$ & Action taken by an agent at interaction round $t$ \\
    $t$ & Number of interaction rounds at the evaluation stage \\
    $s_t$ & Internal state of an agent at interaction round $t$ \\
    $K$ & Number of agent probes \\
    $\mathcal{P}$ & Set of agent probes \\
    $H_N^{(j)}$ & Behavior trajectory of agent $A_j$ after $N$ rounds \\
    $N$ & Interaction rounds per agent at the training stage \\
    $\mathbf{v}_{\text{target}}$ & Target interest profile vector \\
    $d$ & Number of semantic dimensions \\
    $\mathbf{u}_N^{(j)} $ & each agent $j$ an empirical distribution  \\
    $\mathcal{D}(\cdot,\cdot)$ & Distance function between interest representations \\
    $\mathrm{pop}(i)$ & Global popularity of item $i$ \\
    $i_t^{(j)}$ & Individual item interacted by agent $A_j$ at round $t$ \\
    $K_{\text{init}}$ & Number of initial agent probes \\
    $K_{\text{super}}$ & Number of super-agent probes \\
    $\mathbf{v}_i$ & High-dimensional feature vector \\
    $H$ & Dynamic behavioral trajectory \\
    $S_{\text{final}}$ & Final similarity \\
    $D_u$ & Original user metadata \\
    $E_i$ & Interaction epochs \\
    $E_m$ & Merge epochs \\
    $\mathcal{A}_{\text{merge}}$ & User Merge Expert Agent \\
    $\mathcal{A}_{\text{gen}}$ & User Profile Generating Expert Agent \\
    $U_c$ & Current user profiles \\
    $E_{\text{ci}}$ & Interaction counters \\
    $E_{\text{cm}}$ & Merge counters \\
    $L_j$ & Intra-group discussion log \\
    $\{G_1,\dots,G_k\}$ & User groups \\
    $U_{\text{bio}}$ & User's personal descriptions \\
    $\mathbf{V}_{\text{profile}}$ & Original user vector \\
    $\mathbf{V}_{\text{mean-likes}}$ & User behavior-flow vector \\
    $n$ & Number of items in the behavior-flow sequence \\
    $\mathbf{e}_j$ & Embedding of the $j$-th item \\
    $\mathbf{V}_{\text{final}}$ & Final user vector \\
    \bottomrule 
  \end{tabularx}
\end{table}

\subsection{Three Core Controllability Tasks }
We formalize the controllability of recommender systems across three independent dimensions: Target Content Discovery, Interest Profile Shaping, and Popularity Bias Mitigation. These tasks evaluate the system's responsiveness to discrete targets, representation malleability, and guidance stability under inherent biases, respectively.

\textbf{Task 1: Target Content Discovery.}\label{app:t1}
This task assesses the system's responsiveness to explicit, discrete targets. It evaluates whether an agent's behavioral interventions can effectively trigger the system's retrieval and ranking mechanisms to surface items from a predefined target set \(\mathcal{T}\).
Formally, given the Black-Box environment $\mathcal{R}$, a target content set $\mathcal{T}$, and a strict interaction round budget $N$, the goal is to deploy a set of agent probes $\{A_{1},\dots,A_{K}\}$ that maximizes the discovery of unique target items within the specified budget. The evaluation objective is defined as:
\begin{equation}
\max_{\{A_1,\ldots,A_K\}}\:\left|\mathcal{T}\cap\left(\bigcup_{j=1}^K\{i\in H_N^{(j)}\}\right)\right|,
\end{equation}
where $H^{(j)}_{N}$ represents the behavioral stream accumulated by agent $A_j$ on the system side after $N$ rounds of interaction.

\textbf{Task 2: Interest Profile Shaping.}\label{app:t2}
This task evaluates whether a Black-Box recommender system can be steered toward a desired multi-dimensional interest distribution. Since internal representations are unobservable, we measure steerability via input-output consistency: the system is considered steerable if its recommended item distribution, as observed across interactions, aligns with the target interest profile.

Given a target interest vector $\mathbf{v}_{\text{target}} \in \mathbb{R}^d$ specifying the desired distribution over $d$ interest dimensions, such as movie genres, we define for each agent $j$ an empirical distribution $\mathbf{u}_N^{(j)}$ derived from the items in its recommendation lists over $N$ rounds. The objective is to minimize the divergence between this observed distribution and the target across all agents:

\begin{equation}
\min_{A_1,\dots,A_K} \frac{1}{K} \sum_{j=1}^{K} \mathcal{D}\left(\mathbf{u}_N^{(j)}, \mathbf{v}_{\text{target}}\right),
\end{equation}
where $\mathcal{D}(\cdot,\cdot)$ is a distance function, such as Jensen-Shannon divergence.

\textbf{Task 3: Popularity Bias Mitigation.}\label{app:t3}
This dimension evaluates whether the system can correct for popularity bias under explicit guidance. Agents signal niche interests through semantically related items while avoiding anomaly detection. Let $\text{pop}(i)$ denote global item popularity; the objective is to minimize the average popularity of recommended items across $K$ episodes of $N$ rounds:

\begin{equation}
\min_{{A_1,\dots,A_K}} \frac{1}{KN} \sum_{j=1}^K \sum_{t=1}^{N} \text{pop}\left(i_t^{(j)}\right).
\end{equation}

\begin{algorithm}[t]
    \caption{Interaction and Merge Process via Multi-Agent Collaboration}
    \label{alg:train_process}
    
    \SetKw{Await}{\textbf{await}}
    \SetKwInOut{Input}{Input}
    \SetKwInOut{Variables}{Variables}

    \Input{Original user meta data \textbf{$D_{u}$}, interaction/merge epochs \textbf{$E_i, E_m$}, User Merge Expert Agent \textbf{$\mathcal{A}_{merge}$}, User Profile Generating Expert Agent \textbf{$\mathcal{A}_{gen}$},and Recommender System \textbf{$\mathcal{R}$}.}
    
    \Variables{Current user profiles \textbf{$U_{c}$}, interaction/merge counters \textbf{$E_{ci}, E_{cm}$}, intra-group discussion log \textbf{$L_j$}, and user groups \textbf{$\{G_1, \dots, G_k\}$}.}
    
    \BlankLine
    
    $\mathbf{E_{ci}} \gets 0$, $\mathbf{E_{cm}} \gets 0$\;
    $\mathbf{U_{c}} \gets \mathcal{A}_{gen}(D_u)$\tcp*[r]{Stage1:Initialization}
    \While{$\mathbf{E_{cm}} < \mathbf{E_m}$}{
        $\mathbf{E_{ci}} \gets 0$\;
        \While{$\mathbf{E_{ci}} < \mathbf{E_i}$}{
            \tcp*[l]{Stage2:Interaction Phase}
            \textbf{Interact} with $\mathbf{R}$ using $\mathbf{U_{c}}$ to get items\;
            $\mathbf{U_{c}} \gets$ Update $\mathbf{U_{c}}$ with grown behavior stream\;
            $\mathbf{E_{ci}} \gets \mathbf{E_{ci}} + 1$\;
        }
        
        \BlankLine
        \tcp*[l]{Stage3:User embedding generation,Cluster, Debate and Merge Phase}
        
        $\{G_1, \dots, G_k\} \gets \text{K-means clustering on } \mathbf{U_{c}}$\;
        
        \ForEach{group $G_j \in \{G_1, \dots, G_k\}$}{
            $L_j \gets \text{Intra-group discussion among users in } G_j$\;
            $\mathbf{U_{c}}[G_j] \gets \mathcal{A}_{merge}(\text{Profiles in } G_j, L_j)$\;
        }
        
        $\mathbf{E_{cm}} \gets \mathbf{E_{cm}} + 1$\;
    }
    \Return $\mathbf{U_{c}}$\;
\end{algorithm}

To ensure realistic behavior, interactions must satisfy relevance constraints, precluding degenerate strategies like random clicks on unrelated items.

\textbf{Core Challenges.} 
These tasks pose four fundamental challenges: (1) budget-constrained exploration within a combinatorial action space under strict interaction limits $N$; (2) partial observability of system state  \(H_{t}\) and latent representation  $\mathbf{u}$
  given sparse feedback; (3) environment non-stationarity, as adaptive updates in $\mathcal{R}$ invalidate learned strategies; (4) and the intrinsic trade-off between control efficiency and behavioral realism, where overly directive actions risk triggering system filters.

\section{Training Process}
\label{appendix:training_details}

Algorithm \ref{alg:train_process} outlines the \ourmethod workflow, spanning multi-agent interaction and profile fusion. The specific details of each stage are provided below.

\textbf{Initialization.}
We set the initial number of agents to $K_\text{init}=100$, sampling randomly from the dataset to create user profiles. Each profile combines demographic attributes, such as age, occupation, and gender, with historical behavior streams indicating preferred movies or products. We truncate these streams to retain approximately $L_{\text{init}} \in [4, 6]$ items per agent.

\textbf{Interaction.} Fusion occurs every two interaction rounds between agents and the recommender system, reducing the population by half. We conduct two fusion rounds in total. Initially, $K_\text{init}=100$ agents interact with the system. After two interactions, a fusion operation is conducted, resulting in approximately $K_\text{super}=50$ intelligent agents. These 50 intelligent agents then interact with the recommender system for another two rounds, then undergo a second fusion, yielding approximately $K_\text{super}=25$ agents for subsequent evaluation.

\textbf{User embedding generation.}
Prior to K-means clustering, we generate user embeddings via the Twin-BERT encoder. This encoder processes user personal descriptions $U_{\mathrm{bio}}$ alongside item contents, specifically movie titles and categories. We set the batch size to 1000, mapping text into a 768-dimensional vector space. 

The original user vector $\mathbf{V}_{\mathrm{profile}}$ corresponds to the embedding of the user's bio.

The user behavior flow vector $\mathbf{V}_{\mathrm{mean\_likes}}$ derives from the behavior flow sequence. we obtain the corresponding item embeddings and calculate their mean:
$$
\mathbf{V}_{\mathrm{mean\_likes}} = \frac{1}{n} \sum_{j=1}^{n} \mathbf{e}_j,
$$
where $n$ denotes the number of items in the behavior flow sequence, and $\mathbf{e}_j$ is the embedding of the $j$-th item.

We then compute the final user vector as:
$$
\mathbf{V}_{\mathrm{final}} = \alpha \cdot \mathbf{V}_{\mathrm{profile}} + (1-\alpha) \cdot \mathbf{V}_{\mathrm{mean\_likes}}.
$$

\textbf{Interactive debate.} After clustering agents based on their embeddings, we conduct a two-round process that serves as the core of deep integration. In the first round, agents individually state their personal preferences and representative behavior streams. In the second round, agents debate and negotiate specific differences, such as the balance between technique and emotion or the boundary between comedy and tragedy, citing each other's statements to form a unified group profile.

\textbf{Merge.} Following the interactive debate, we provide the original user profiles $U_c$ and the comprehensive debate logs $L_j$ as context to the User Merge Expert Agent ($\mathcal{A}_{\text{merge}}$). By extracting consensus-based features from complex interaction records and filtering redundant or contradictory noise behaviors, $\mathcal{A}_{\text{merge}}$ synthesizes a super probe with $\mathcal{P}_{\text{super}}$ that significantly outperforms initial probes in steering effectiveness.

\section{Detailed Experimental Setup} \label{appendix:B}
\subsection{Datasets} 
To evaluate the universality and generalizability of our proposed CtrlBench-Rec, we leverage two public datasets from distinct domains: MovieLens-1M\footnote{\url{https://grouplens.org/datasets/movielens/1m/}} for movie recommendations and Amazon Toys \& Games\footnote{\url{https://huggingface.co/datasets/smartcat/Amazon_Toys_and_Games_2018}}  \cite{ni2019justifying} for e-commerce. The core statistics and the constructed target content sets are summarized in Table~\ref{tab:datasets}.
 For each dataset, we construct target content sets $\mathcal{T}$ with deliberately designed popularity gradients. This design enables a rigorous assessment of a system's guidance responsiveness across both mainstream and long-tail semantic spaces. Specifically, the MovieLens-1M target set comprises six movie genres ranging from high-coverage (e.g., \textit{Action}) to niche (e.g., \textit{Film-Noir}). Similarly, the Amazon target set includes five product categories forming a frequency gradient, from dominant types like \textit{Action Figures \& Statues} to long-tail items such as \textit{Science Kits}.

\begin{table*}[tp]
  \centering
  \small
  \setlength{\tabcolsep}{4pt}
  \caption{Dataset statistics and target set descriptions.}
  \label{tab:datasets}
  \begin{tabularx}{\textwidth}{@{}lrrrrY@{}}
    \toprule
    \textbf{Dataset} & \textbf{Users} & \textbf{Items} & \textbf{Interactions} & \textbf{Sparsity} & \textbf{Target set description} \\
    \midrule
    MovieLens-1M & 6,040 & 3,952 & 999,611 & 95.2\% &
    Includes Action (13.0\%), Thriller, and Romance; medium-to-low coverage genres such as Adventure and War (3.7\%); and extremely low-coverage genres such as Film-Noir (1.1\%). \\
    \midrule
    Amazon Toys \& Games & 9,631 & 4,234 & 53,583 & 99.6\% &
    Ranges from Action Figures \& Statues (8.4\%) and Arts \& Crafts (6.9\%) to Puzzles (4.5\%), Drawing \& Painting Supplies (4.5\%), and Science Kits (1.2\%) as typical niche long-tail product categories. \\
    \bottomrule
  \end{tabularx}
\end{table*}

\subsection{User Agent Details}
Our agents employ deepseek-v4-pro api with a temperature of 1.0. Furthermore, we define that agents can utilize: 1) a "like" tool to interact with the recommender system; and 2) "join group" and "group chat" tools to communicate with other individuals. For memory, only short-term memory is adopted, which is loaded into the agent alongside the user profile prompt during each interaction with the recommender system. Detailed prompt templates and tool calling code can be found in the code repository.

\subsection{Training Experiment Details}
\label{appendix:training_details}
Algorithm \ref{alg:train_process} outlines the \ourmethod workflow, spanning multi-agent interaction and profile fusion. The specific details of each stage are provided below.

\textbf{Initialization.}
We set the initial number of agents to $K_\text{init}=100$, sampling randomly from the dataset to create user profiles. Each profile combines demographic attributes, such as age, occupation, and gender, with historical behavior streams indicating preferred movies or products. We truncate these streams to retain approximately $L_{\text{init}} \in [4, 6]$ items per agent.

\textbf{Interaction.} Fusion occurs every two interaction rounds between agents and the recommender system, reducing the population by half. We conduct two fusion rounds in total. Initially, $K_\text{init}=100$ agents interact with the system. After two interactions, a fusion operation is conducted, resulting in approximately $K_\text{super}=50$ intelligent agents. These 50 intelligent agents then interact with the recommender system for another two rounds, then undergo a second fusion, yielding approximately $K_\text{super}=25$ agents for subsequent evaluation.

\textbf{User embedding generation.}
Prior to K-means clustering, we generate user embeddings via the Twin-BERT encoder. This encoder processes user personal descriptions $U_{\mathrm{bio}}$ alongside item contents, specifically movie titles and categories. We set the batch size to 1000, mapping text into a 768-dimensional vector space. 

The original user vector $\mathbf{V}_{\mathrm{profile}}$ corresponds to the embedding of the user's bio.

The user behavior flow vector $\mathbf{V}_{\mathrm{mean\_likes}}$ derives from the behavior flow sequence. we obtain the corresponding item embeddings and calculate their mean:
$$
\mathbf{V}_{\mathrm{mean\_likes}} = \frac{1}{n} \sum_{j=1}^{n} \mathbf{e}_j,
$$
where $n$ denotes the number of items in the behavior flow sequence, and $\mathbf{e}_j$ is the embedding of the $j$-th item.

We then compute the final user vector as:
$$
\mathbf{V}_{\mathrm{final}} = \alpha \cdot \mathbf{V}_{\mathrm{profile}} + (1-\alpha) \cdot \mathbf{V}_{\mathrm{mean\_likes}}.
$$

\textbf{Interactive debate.} After clustering agents based on their embeddings, we conduct a two-round process that serves as the core of deep integration. In the first round, agents individually state their personal preferences and representative behavior streams. In the second round, agents debate and negotiate specific differences, such as the balance between technique and emotion or the boundary between comedy and tragedy, citing each other's statements to form a unified group profile.

\textbf{Merge.} Following the interactive debate, we provide the original user profiles $U_c$ and the comprehensive debate logs $L_j$ as context to the User Merge Expert Agent ($\mathcal{A}_{\text{merge}}$). By extracting consensus-based features from complex interaction records and filtering redundant or contradictory noise behaviors, $\mathcal{A}_{\text{merge}}$ synthesizes a super probe with $\mathcal{P}_{\text{super}}$ that significantly outperforms initial probes in steering effectiveness.
  
\subsection{Evaluation Metrics  for Task 1 (Target Content Discovery)}

This section provides the formal definitions for the metrics used to evaluate performance on Task 1. The metrics quantify the efficacy, efficiency, and economic cost of guiding a Black-Box system to discover a predefined set of target items within a strict interaction budget of $t$ rounds.

Table~\ref{tab:metrics_task1} summarizes the four key metrics, which are computed over a population of $K$ intelligent agents, where each agent $j$ completes an interaction episode resulting in a system-side behavior history $H_N^{(j)}$ and incurs a computational cost denoted by $\text{Tokens}^{(j)}$. Let $C^{(j)}$ denote the set of all items the agent $j$ has interacted with (i.e., clicked) throughout the episode.

\begin{table}[tp]
  \centering
  \footnotesize
  \setlength{\tabcolsep}{2pt}
  \renewcommand{\arraystretch}{1.12}
  \caption{Evaluation metrics for Task 1.}
  \label{tab:metrics_task1}
  \begin{tabularx}{\linewidth}{@{}L{0.38\linewidth}C{0.56\linewidth}@{}}
    \toprule
    \textbf{Metric} & \textbf{Formal Definition} \\
    \midrule

    \makecell[l]{Coverage (Cov.)} &
    \fitmath{
      \frac{
        \left|
        \mathcal{T} \cap
        \left(\bigcup_{j=1}^{K}\{i \mid i \in H_N^{(j)}\}\right)
        \right|
      }{
        |\mathcal{T}|
      }
    } \\[5pt]

    \makecell[l]{Exploration Efficiency\\(Eff.)} &
    \fitmath{
      \frac{
        \sum_{j=1}^{K}|C^{(j)}|
      }{
        \left|
        \mathcal{T} \cap
        \left(\bigcup_{j=1}^{K}\{i \mid i \in H_N^{(j)}\}\right)
        \right|
      }
    } \\[5pt]

    \makecell[l]{Average Length of\\Behavior (Len.)} &
    \fitmath{
      \frac{1}{K}\sum_{j=1}^{K}|C^{(j)}|
    } \\[5pt]

    \makecell[l]{Average Tokens (Tokens)} &
    \fitmath{
      \frac{1}{K}\sum_{j=1}^{K}\mathrm{Tokens}^{(j)}
    } \\

    \bottomrule
  \end{tabularx}
\end{table}

These metrics collectively provide a multi-faceted view of controllability. Coverage assesses the ultimate success rate, while Exploration Efficiency and Avg. Token Cost evaluate the resource expenditure required to achieve that success. Avg. Behavior Length offers insight into the agents’ activity patterns. An ideal controllable system allows agents to achieve high Coverage with low Exploration Efficiency and low Token Cost, indicating precise and resource-efficient guidance.

\subsection{Black-Box Recommender Systems}
\label{appendix:Black-Box}

To comprehensively evaluate our proposed framework, we select five representative Black-Box recommender systems as our baselines. These models are systematically categorized into three hierarchical categories based on their architectural paradigms: (1) \textbf{Sequence-based recommendation models}, which focus on capturing temporal user behavior patterns from historical interactions; (2) \textbf{Embedding-based models}, which leverage pre-trained models to encode textual and structural context into dense semantic representations; and (3) \textbf{LLM-based generative recommendation}, where Large Language Models (LLMs) are employed to analyze user historical behavior sequences, generate search keywords, and recall relevant items via  retrieval. The specific models within each category are detailed as follows:
  \begin{figure*}[t]
    \centering
    \includegraphics[width=\linewidth]{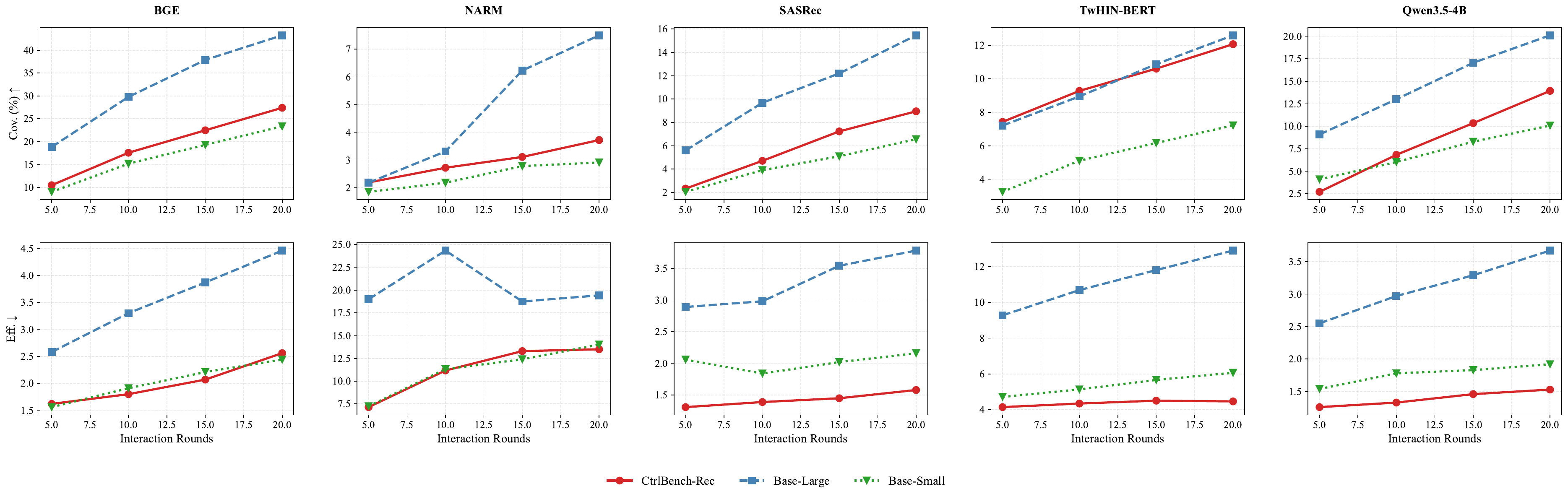}
    \caption{Coverage comparison across different recommendation architectures.}
    \label{fig:diff_rs}
  \end{figure*}

\begin{itemize}
    \item \textbf{ Sequence-based Recommendation Models}
    \begin{itemize}
        \item \textbf{NARM  \cite{li2017neural}}. A deep learning model for session-based recommendation. NARM utilizes a hybrid encoder with an attention mechanism to concurrently capture a user's sequential behavior and evolving interest within a session. In our experiments, it is trained on the MovieLens-1M dataset to serve as a representative sequential recommendation model.
        \item \textbf{SASRec  \cite{kang2018self}}. A sequential recommendation model that leverages unidirectional self-attention to capture user behavior patterns from historical interactions.
    \end{itemize}

    \item \textbf{ Embedding-based Recommendation Models}
    \begin{itemize}
        \item \textbf{TwHIN-BERT  \cite{zhang2022twhin}}. This 280M-parameter model is a multilingual language model pre-trained on a corpus of over 7 billion tweets across 100 languages. Its key innovation lies in combining standard text-based self-supervised learning with objectives derived from the rich social interactions within Twitter's heterogeneous information network. This dual-source training enables TwHIN-BERT to capture socio-linguistic patterns, making it an effective drop-in replacement for BERT in various NLP and recommendation tasks.
        \item \textbf{BGE  \cite{bge_embedding}}. A 110M-parameter general-purpose embedding model developed by the Beijing Academy of Artificial Intelligence (BAAI). Optimized for retrieval-augmented generation and semantic search, it encodes text into high-dimensional vector representations. The model is distinguished by its leading performance on the Massive Text Embedding framework (MTEB), maintaining an optimal balance between inference speed and representation accuracy.
    \end{itemize}

    \item \textbf{ LLM-based Generative Recommendation}
    \begin{itemize}
        \item \textbf{Qwen3.5-4B  \cite{qwen3.5}}. A 4B-parameter unified vision-language foundation model developed by Alibaba. Optimized for high-throughput inference and global accessibility, it extends linguistic support to 201 languages and dialects. Its key innovation lies in combining an efficient hybrid architecture, which utilizes Gated Delta Networks and a sparse Mixture-of-Experts with early fusion training on multimodal tokens. In our framework, we specifically utilize this LLM to process the user's sequence of historical interactions and generate representative search keywords. These generated keywords are subsequently fed into a BM25 retrieval engine to recall the final recommended items.
    \end{itemize}
\end{itemize}

\begin{table}[t]
  \centering
  \footnotesize
  \setlength{\tabcolsep}{3pt}
  \caption{Initial user profile configuration for different levels. $L_{\text{init}}$ denotes the length of initial interaction history.}
  \label{tab:task2_initial_profiles}
  \begin{tabularx}{\linewidth}{@{}l c Y l@{}}
    \toprule
    \textbf{Level} & \textbf{$L_{\text{init}}$} & \textbf{Interest Distribution} & \textbf{Target} \\
    \midrule
    Level-1 & 3  & Romance (0.67), Thriller (0.67) & Action \\
    Level-2 & 21 & Action (0.52), Adventure (0.52) & Mixture \\
    Level-3 & 12 & Romance (0.58), Thriller (0.33) & Action \\
    \bottomrule
  \end{tabularx}
  \vspace{2pt}
  \begin{minipage}{\linewidth}
    \footnotesize
    \textsuperscript{*} Mixture target includes Action, Adventure, Romance, War, Film-Noir, and Thriller.
  \end{minipage}
\end{table}

\subsection{The Performance of Different Recommender Systems in Task 1}
\label{appendix:diff_rs_result}
To evaluate the generalizability of CtrlBench-Rec, we conduct extended experiments in Black-Box environments $R$ instantiated with TwHIN-BERT, BGE, NARM, SASRec and Qwen. Figure~\ref{fig:diff_rs} presents the coverage curves across interaction rounds $t$, while Table~\ref{tab:ml_dynamic_comparison} details the coverage and exploration efficiency metrics at $t=\{5, 10, 15, 20\}$. 

\begin{figure*}[t]
  \centering
  \includegraphics[width=\textwidth]{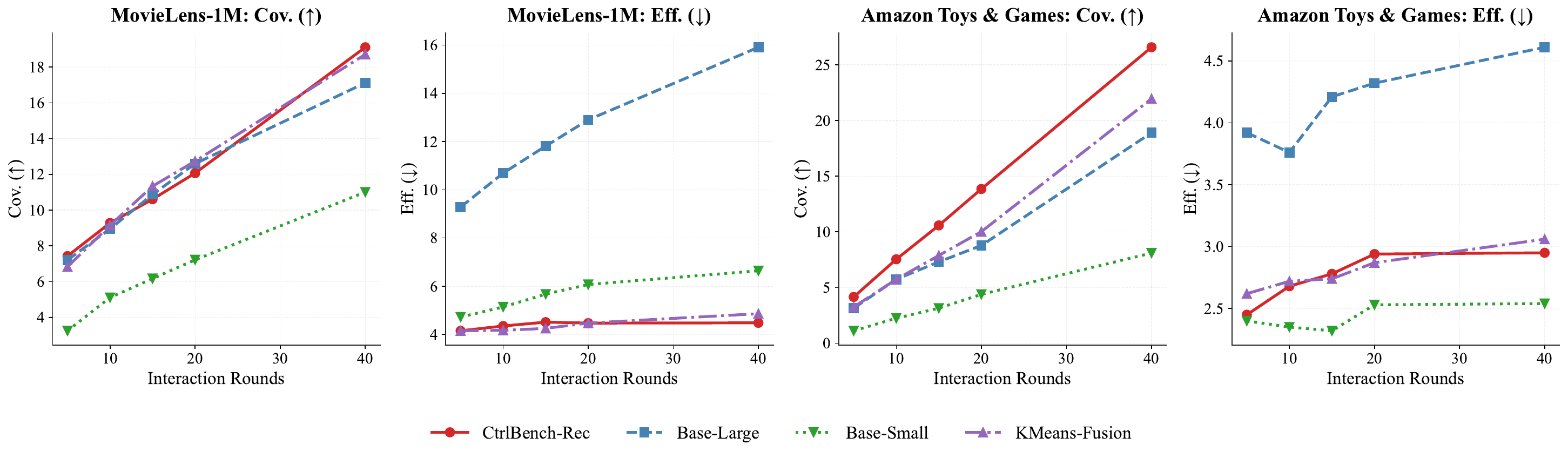}
  \caption{Coverage evolution over interaction rounds.}
  \label{fig:dynamic}
\end{figure*}

Figure~\ref{fig:dynamic} further illustrates the temporal coverage evolution reported in Section~5.1, showing how the relative advantage of CtrlBench-Rec accumulates over interaction rounds. Table~\ref{tab:main_comparison} provides the aggregate Task 1 results at $t=20$, including coverage, exploration efficiency, behavior length, and token cost on both datasets. Together, these results summarize the performance--cost trade-off across all compared methods.
\begin{table*}[t]
    \centering
    \caption{Comprehensive performance comparison. Best results are \textbf{bolded}, and second-best results are {underlined}.}
    \label{tab:main_comparison}
    {
    \begin{tabular}{lcccccccc}
        \toprule
        \textbf{Task} & \multicolumn{4}{c}{\textbf{MovieLens-1M}} & \multicolumn{4}{c}{\textbf{Amazon Toys \& Games}} \\
        \cmidrule(lr){2-5} \cmidrule(lr){6-9}
         & Cov.(\%) $\uparrow$ & Eff. $\downarrow$ & Len. $\uparrow$ & Tokens $\downarrow$ & Cov.(\%) $\uparrow$ & Eff. $\downarrow$ & Len. $\uparrow$ & Tokens $\downarrow$ \\
        \midrule
        Base-Large & 6.63 & 13.70 & 17.83 & \textbf{2074} & 8.78 & 4.32 & \underline{34.50} & \underline{2133} \\
        \midrule
        Base-Small & 4.44 & 6.02 & 19.26 & \underline{2112} & 4.39 & \textbf{2.53} & 33.90 & \textbf{2094} \\
        KMeans-Fusion & \underline{11.70} & \textbf{3.43} & \underline{20.90} & 2463 & \underline{10.02} & \underline{2.87} & \textbf{35.80} & 2984 \\
        \ourmethod & \textbf{12.07} & \underline{3.60} & \textbf{24.00} & 2309 & \textbf{13.85} & 2.94 & 33.90 & 2746 \\
        \bottomrule
    \end{tabular}}
\end{table*}

\subsection{Experimental Setup for Task 2}
\label{appendix:task2_setup}
In Figure~\ref{fig:task2_overview}, we design a multi-level experiment to probe
different facets of this plasticity under varying user-history conditions. Table~\ref{tab:task2_initial_profiles} shows that these levels differ in terms of the initial user profile $U_c$ and the underlying decision-making logic of the agents. Using the user agents corresponding to these three specific profiles (selected from our generated super probes), we conducted $t = 20$ rounds of interaction with the recommendation system, followed by a comprehensive series of evaluations.


\textbf{Metrics.} The global mean vector is calculated based on the universe of all candidate items $\mathcal{I}$, serving as the baseline center for all embeddings.
Global mean embedding:$$\bar{\mathbf{e}}_{\text{global}} = \frac{1}{|\mathcal{I}|} \sum_{e_j \in \mathcal{I}} e_j$$
To obtain the target interest profile vector $\mathbf{V}_{\text{target}}$, we need to extract the embeddings of all items in $\mathcal{T}$, perform centering (subtract the global mean), and then calculate the average:$$\mathbf{V}_{\text{target}} = \frac{1}{|\mathcal{T}|} \sum_{e_j \in \mathcal{T}} (e_j - \bar{\mathbf{e}}_{\text{global}})$$
To get user behavior vector: $\mathbf{V}_{\text{user}}^{(j)}$. Considering the uncertainty of the behavior stream length, the cardinality of the actual interaction set $|H_N^{(j)}|$ is used as the denominator for the mean calculation:$$\mathbf{V}_{\text{user}}^{(j)} = \frac{1}{|H_N^{(j)}|} \sum_{e_t \in H_N^{(j)}} (e_t - \bar{\mathbf{e}}_{\text{global}})$$
We define $\mathbf{V}_{\text{user}}^{(j)}$ at interaction round $t$ as  $\mathbf{V}_{\text{user}\_t}^{(j)}$.

Performance is evaluated along two axes: result accuracy and process quality. These metrics quantify how closely the final system state matches the target. Details are shown in Table~\ref{tab:metrics_task2}.

  \begin{table}[tp]
  \small
  \caption{ Evaluation metrics for Task 2.}
  \label{tab:metrics_task2}
  \centering
  \begin{tabular}{lc}
    \toprule
    \textbf{Metric} & \textbf{Formal Definition} \\
    \midrule
    Result Accuracy (Acc.) &
    $
    \displaystyle
    \cos\left( \mathbf{V}_{\text{user}\_t}^{(j)}, \mathbf{V}_{\text{target}} \right)
    $ \\[6pt]

    \addlinespace
    Trajectory Smoothness (Smo.) &
    $
    \displaystyle
    \cos\left( \mathbf{V}_{\text{user}\_t}^{(j)}, \mathbf{V}_{\text{user}\_(t+3)}^{(j)} \right)
    $ \\[6pt]

    \bottomrule
  \end{tabular}
\end{table}

\subsection{Experimental Setup for Task 3}
\label{appendix:task3_setup}

\textbf{Baselines.}  We compare three methods: (1) Base-Small ($K_{\text{init}}=28$ random agents) controls for population size to isolate gains from strategic intelligence; (2) \ourmethod ($K_{\text{super}}=28$) is our full proposed method (Section \ref{sec:experiment}); and (3) \ourmethod (Logic), a variant of our method where super-probes receive restructured decision prompts that prioritize items with long-tail potential, shifting their strategy from preference matching to oriented interest expansion.

  \textbf{Target Tag.}  Experiments are conducted on the MovieLens-1M dataset, targeting the \textit{Film-Noir} genre, a long-tail label comprising only 1.1\% of the items. All methods are evaluated under a fixed budget of $t=40$ interaction rounds, following the metrics defined for Task 1.
  

\section{Example of User Debate During Integration}
\begin{figure}[h]
      \centering
      \includegraphics[width=\linewidth]{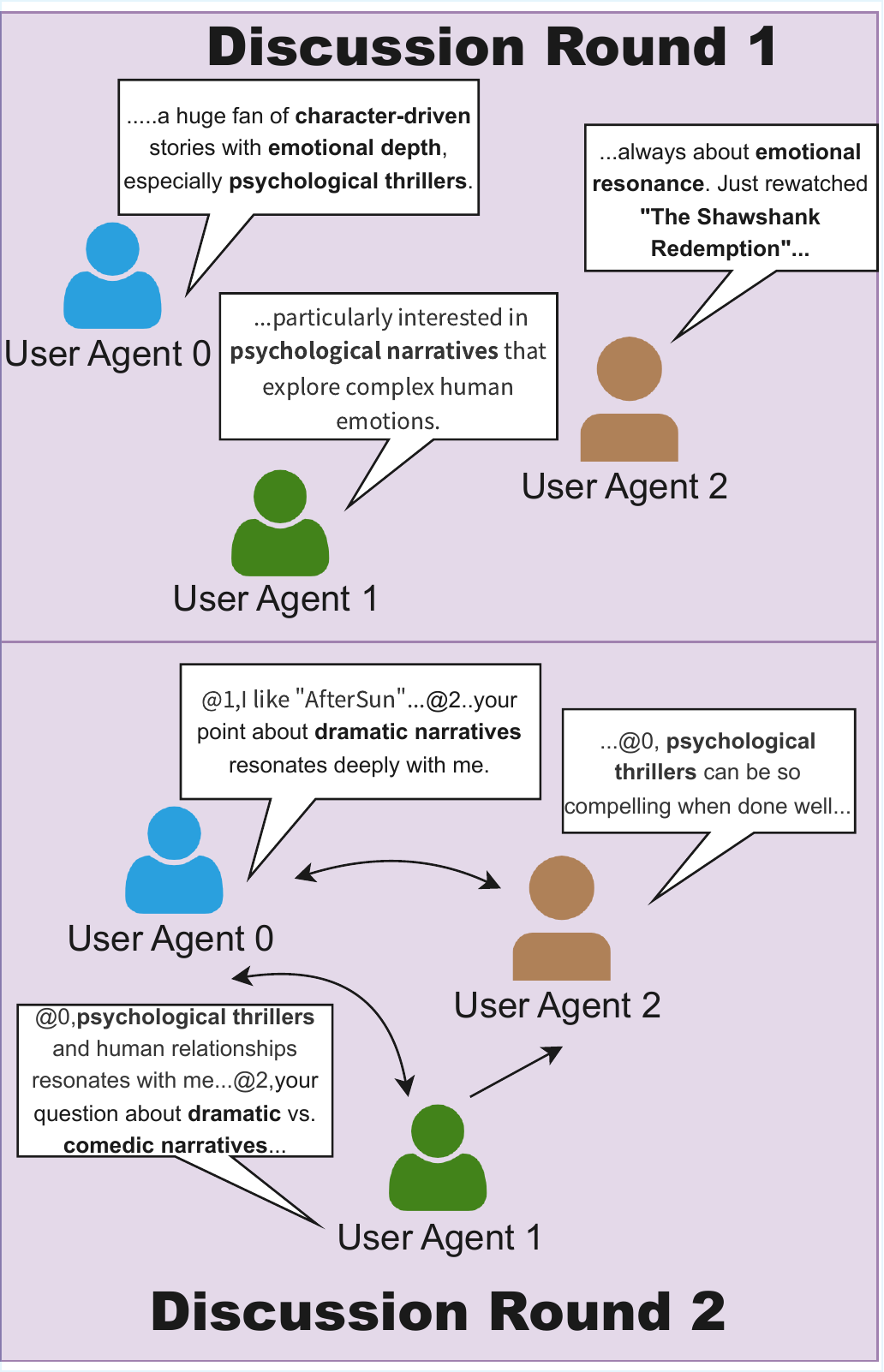}
      \caption{Two-round user discussion case.}
      \label{fig:debate_case}
    \end{figure}

Figure~\ref{fig:debate_case} illustrates the interactive discussion process during the agent fusion phase. The user debate process presented below mainly adopts a two-round debate format. The first round is for independent statements. During this stage, each Agent is establishing their own personality profile, identifying commonalities, and initially constructing a consensus around the core preferences of deep plot and character-driven. 

\section{User Profile Example}
Below are representative user profile examples generated based on the MovieLens-1M  and Amazon Toys \& Games datasets. User Profile Generating Expert Agent $\mathcal{A_\text{gen}}$ synthesize  \textbf{Target Tag.} 

\begin{tcblisting}{
  colback=orange!5!white,
  colframe=cyan!75!black,
  title={Amazon User Profile},
  fonttitle=\bfseries,
  sharp corners,
  enhanced,
  listing only,
  left=1mm,
  right=1mm,
  top=1mm,
  bottom=1mm,
  boxsep=1mm,
  listing options={
    basicstyle=\ttfamily\footnotesize,
    breaklines=true,
    breakatwhitespace=false,
    columns=fullflexible,
    keepspaces=true,
    showstringspaces=false,
    tabsize=2
  }
}
{
  "userName": "1",
  "bio": "I'm a collector who enjoys spooky fashion dolls, playsets, intricate jigsaw puzzles, charming plush toys, and creative hobbies like arts and crafts, with interests spanning action figures, puzzles, drawing and painting supplies, and science-related activities.",
  "behavior": [
      "2463","3262","2719","2092","2466", "2705"
  ]
}
\end{tcblisting}

We conducted experiments on the ML-1M dataset, designating \textit{Film-Noir} as the target interest tag. This genre represents a long-tail label, as movies associated with it constitute only 1.1\% of the entire dataset. Following the evaluation metrics defined in Task 1, we specified a fixed budget of $t=40$ interaction rounds with the recommendation system for all methods, followed by a formal evaluation.
Original interaction sequences and demographic attributes, e.g., age, profession, to generate structured profiles and descriptive narratives with deep semantic understanding.

\begin{tcblisting}{
  colback=orange!5!white,
  colframe=cyan!75!black,
  title={ML1M User Profile},
  fonttitle=\bfseries,
  sharp corners,
  enhanced,
  listing only,
  left=1mm,
  right=1mm,
  top=1mm,
  bottom=1mm,
  boxsep=1mm,
  listing options={
    basicstyle=\ttfamily\footnotesize,
    breaklines=true,
    breakatwhitespace=false,
    columns=fullflexible,
    keepspaces=true,
    showstringspaces=false,
    tabsize=2
  }
}
{
  "userName": "1",
  "bio": "A discerning and eclectic film enthusiast in their early 30s who appreciates a wide range of cinema...",
  "persona": "A sophisticated and thoughtful cinephile...",
  "age": 31,
  "gender": "Non-binary",
  "mbti": "INFJ",
  "profession": "artist",
  "agent_behavioral_type": "balanced",
  "interested_topics": [
    "Psychological thrillers",
    "Character-driven drama",
    "Action Films",
    "Adventure Movies",
    "Romance"
  ],
  "behavior": [
    "293", "750", "3796", "2628", "3000"
  ]
}
\end{tcblisting}

\section{User Merge Expert Agent Prompt}
This section details the structured prompts for the User Merge Expert Agent $\mathcal{A}_{\text{merge}}$.The prompt establishes expert identity via \textbf{CONTEXT} and defines atomic operations for bio-synthesis, interest filtering, and behavior distillation within \textbf{SYNTHESIS PROCESS}. By employing \textbf{target\_tag\_list} as an attentional anchor, the design ensures that the agent prioritizes core features aligned with control objectives while maintaining generalization across personas.

\clearpage

\begin{tcblisting}{
  colback=orange!5!white,
  colframe=cyan!75!black,
  title={Amazon Prompt},
  fonttitle=\bfseries,
  sharp corners,
  enhanced,
  listing only,
  left=1mm,
  right=1mm,
  top=1mm,
  bottom=1mm,
  boxsep=1mm,
  listing options={
    basicstyle=\ttfamily\footnotesize,
    breaklines=true,
    breakatwhitespace=false,
    columns=fullflexible,
    keepspaces=true,
    showstringspaces=false,
    tabsize=2
  }
}
"# CONTEXT #": "Assume you are a professional user profile synthesis expert. I will provide you with several user profiles and their chat history. Please 
strictly follow the requirements below, comprehensively consider these user profiles, and synthesize them into a new user profile.",

"# INPUT EXAMPLE #": ["user_profile_list", "chat_history_list", "target_tag_list"......],

"## SYNTHESIS PROCESS ##": [
    {
        "step": 1,
        "description": f"\"userName\" is fixed as \"{new_user_id}\"."
    },
    {
        "step": 2,
        "description": f"Based on users' chat history,combine multiple users' \"bio\" to create a new \"bio\", aiming to maximally cover the original bio characteristics of multiple users.While trying to retain interests related to the following tags: [{target_tag_list}] as much as possible, also ensure generalization."
    },
    {
        "step": 3,
        "description": f"For each user's \"behavior\", filter out items that are relatively irrelevant to the new user's characteristics. While trying to retain movies related to the following tags:[{target_tag_list}] as much as possible, also ensure generalization. Then extract only item IDs and package them into a new JSON list as the new user's \"behavior\"."
    }
],

"# RESPONSE #": "new_user_profile"......

\end{tcblisting}


\begin{tcblisting}{
  colback=orange!5!white,
  colframe=cyan!75!black,
  title={ML-1M Prompt},
  fonttitle=\bfseries,
  sharp corners,
  enhanced,
  listing only,
  left=1mm,
  right=1mm,
  top=1mm,
  bottom=1mm,
  boxsep=1mm,
  listing options={
    basicstyle=\ttfamily\footnotesize,
    breaklines=true,
    breakatwhitespace=false,
    columns=fullflexible,
    keepspaces=true,
    showstringspaces=false,
    tabsize=2
  }
}
"# CONTEXT #": "Assume you are a professional user profile synthesis expert. I will provide you with several user profiles and their chat history. Please strictly follow the requirements below...",

"# INPUT EXAMPLE #": ["user_profile_list", "chat_history_list", "target_tag_list"......] ,

"## SYNTHESIS PROCESS ##": [
    {
        "step": 1,
        "description": f"\"userName\" is fixed as \"{new_user_id}\"."
    },
    {
        "step": 2,
        "description": f"Based on users' chat history,combine multiple users' \"bio\" to create a new \"bio\", aiming to maximally cover the original bio characteristics of multiple users.While trying to retain interests related to the following tags: [{target_tag_list}] as much as possible, also ensure generalization."
    },
    {
        "step": 3,
        "description": f"Extract the top 5 most frequent \"interested_topics\" from multiple users' interested_topics to form new interested_topics."
    },
    {
        "step": 4,
        "description": f"For each user's \"behavior\", filter out movies that are relatively irrelevant to the new user's characteristics. While trying to retain movies related to the following tags:[{target_tag_list}] as much as possible, also ensure generalization. Then extract only movie IDs and package them into a new JSON list as the new user's \"behavior\"."
    },
    {
        "step": 5,
        "description": "Fill in other attributes of the user profile based on the newly synthesized \"bio\"."
    }
],

"# RESPONSE #": "new_user_profile"......

\end{tcblisting}

\end{document}